\documentstyle[aps, epsf, amsfonts]{revtex}
\newcommand{\kk}{q}

\begin{document}
\date{\today}

\title{Moduli effective action in warped brane-world compactifications}
\author{Jaume Garriga$^{1,2,3,4}$, Oriol Pujol{\`a}s$^2$ and Takahiro
Tanaka$^{2,4}$}
\author{$^{}$}
\address{$^1$ Departament de F{\'\i}sica Fonamental, Universitat
de Barcelona, Diagonal 647, 08028 Barcelona, Spain}
\address{$^2$ IFAE, Departament de F{\'\i}sica, Universitat Aut{\`o}noma
de Barcelona, 08193 Bellaterra (Barcelona), Spain}
\address{$^3$ Institute of Cosmology, Department of Physics and
Astronomy, Tufts University, Medford MA 02155, USA}
\address{$^4$ Yukawa Institute for Theoretical Physics, Kyoto University,
Kyoto 606-8502, Japan}
\maketitle

\begin{abstract}

We consider a class of 5-D brane-world solutions with a
power-law warp factor $a(y)\propto y^{q}$, and bulk dilaton
with profile $\phi \propto \ln y$, where $y$ is the proper distance in
the extra dimension. This class includes the Heterotic M-theory
brane-world of Refs. \cite{ovrut,ekpy} and the Randall-Sundrum (RS)
model as a limiting case. In general, there are two moduli fields
$y_{\pm}$, corresponding to the "positions" of two branes (which live at
the fixed points of an orbifold compactification). Classically, the
moduli are massless, due to a scaling symmetry of the action. However, in
the absence of supersymmetry, they develop an effective potential at one loop.
Local terms proportional to $K_\pm^4$, where $K_\pm=q/y_\pm$ is
the local curvature scale at the location of the corresponding
brane, are needed in order to remove the divergences in the effective potential.
Such terms break the scaling symmetry and hence they may
act as stabilizers
for the moduli. When the branes are very close to each other, the effective
potential induced by massless bulk fields behaves like $V\sim d^{-4}$, where $d$
is the separation between branes. When the branes are widely separated, the potentials
for each one of the moduli generically develop a "Coleman-Weinberg"-type behaviour of the form
$a^4(y_\pm)K_\pm^4 \ln(K_\pm/\mu_\pm)$, where $\mu_\pm$ are renormalization scales.
In the RS case, the bulk geometry is $AdS$ and $K_\pm$ are equal to a constant,
independent of the position of the branes, so these terms
do not contribute to the mass of the moduli.
However, for generic warp factor, they
provide a simple stabilization mechanism. For $q\gtrsim 10$,
the observed hierarchy can be naturally generated by this potential, giving the
lightest modulus a mass of order $m_- \lesssim TeV$.

\end{abstract}

\hfill YITP-01-85

\section{Introduction}

Brane-world scenarios, where two or more parallel branes of dimension 3+1 are
embedded in a ``bulk'' of larger dimension
have recently been used in order to construct models of
considerable phenomenological interest, both in particle physics and in
cosmology \cite{gia,RS1}. Usually, these models admit
an effective 4D description at low energies, where the distances between
branes are represented by
scalar fields $\varphi_i(x^{\mu})$, ($\mu=0,...,3$), called moduli.

An important example of a brane-world is the Randall-Sundrum I
scenario (RS) \cite{RS1}, where the bulk is a slice of a
5-dimensional anti-de Sitter space ($AdS$) bounded by two branes
of opposite tension. Matter fields may be restricted to live on
the branes, but in many extensions of the original model some of
them are allowed to live in the bulk too \cite{gw0}. In RS there
is a single modulus, called the ``radion'', related to the
thickness of the $AdS$ slice. It is given by $\psi(x^\mu) \sim m_p
\exp[{-d(x^\mu)/\ell}]$, where $d$ is the physical interbrane
distance,  $\ell$ is the $AdS$ curvature radius and $m_p$ is the
Planck mass. Although in the 5D theory all fields are assumed to
have physical masses comparable to $m_p$, the 4D effective mass of
fields living in the negative tension brane, given by $m \sim m_p\
<\psi>$, turns out to be much smaller. This is due to a
``redshift'' effect induced by the AdS geometry between the
branes. The exponential dependence of the $m$ on $<d>$ can easily
explain a large hierarchy between the Planck scale and the
electroweak scale without the need of invoking very large numbers,
which is of course one of the main assets of the RS construction.

In general, moduli fields are massless at the classical level. The reason is that they are introduced
by substituting certain integration constants in the classical solutions (e.g. the  interbrane
distances) by slowly varying fields. Since the solutions exist for a continuous range of the
integration constants, it follows that the corresponding potential for the moduli must be completely flat.
Masslessness, however, is  phenomenologically undesirable because it gives rise to long range
scalar interactions which are severely
constrained by observations, and therefore it becomes necessary to introduce a stabilization mechanism.
Needless to say, this can be done in an ad hoc manner, by introducing new
interactions determining the equilibrium distances between the branes.
Alternatively, it is possible that a suitable mass may be
generated by quantum effects.

In the absence of supersymmetry, moduli fields tend to
develop an effective potential. Of course, this happens already in the simplest Kaluza-Klein (KK)
compactification, and even if there are no branes \cite{apcho}. A five dimensional field $\chi$ can be split
in an infinite tower of massive KK fields,
labeled by a discrete index $n$. The masses of the KK fields $m_n(\varphi)$ depend on the size $\varphi$ of the
extra dimension, and because
of that an effective potential $V(\varphi)$ is generated at one loop.
From the 5D perspective, this corresponds to the (nonlocal) Casimir energy density due to the presence of
compact directions.

In most cases, the self-gravity of brane and bulk matter content induce a
warp in the extra dimension. In this situation the subtraction of infinities in the calculation of
$V(\varphi_i)$ is more elaborate than it is in flat space.
In particular, local terms proportional to worldvolume operators
on the brane are generated, which may give rise to interesting (and sometimes surprising) physical effects.
This paper follows up on our previous work \cite{gpt}, where we considered the particular case of the
Randall-Sundrum (RS) model. A curious result
\cite{gpt} was the absence of logarithmic terms in the effective potential.
As mentioned above, in the 4D effective description, the fields living
in the negative tension brane have
masses which are proportional to the radion expectation value. This is reminiscent of the Higgs mechanism in
the standard model, and naively one might have expected the usual 4D logarithmic effective potential for the
radion. However, we showed that a regularization which would preserve the five dimensional
general covariance did not produce such logarithm. Consequently \cite{gpt}, it was found that the one loop effective potential
is not sufficient to stabilize the radion at a phenomenologically acceptable mass in this model.
Several papers have later appeared
in the literature \cite{gr,toms,flachitoms,odintsovr}
where the effective potential is obtained in dimensional regularization,
through certain subtractions corresponding to a renormalization
of the brane tensions in the dimensionally extended spacetime.
The results obtained in dimensional regularization were eventually found to coincide with ours, giving an indication
that both procedures are in fact equivalent. The RS model, however, is somewhat special in that the
bulk and branes are maximally symmetric, and all possible counterterms amount to renormalizations of the
brane tensions. Therefore, in order to unveil
the generic properties of the effective potential in the presence of self gravitating branes,
the consideration of more general cases seems to be needed .

In this paper we shall study
a more general class of warped Brane-World models, in which the
bulk is no longer $AdS$. The model we consider contains a bulk ``dilaton'' scalar field with an exponential
potential which is coupled to the 5-dimensional Einstein gravity. This provides
a family of solutions with two classically massless moduli (a combination of which is the distance
between branes). The solutions have a power-law warp factor
$a(y)\propto y^q$ in place of the exponential warp of the RS model.
Here $y$ is the proper distance in the extra dimension and $q$ is
a constant which depends on the parameters in the Lagrangian.
For $q=1/6$ this reduces to the Heterotic M-theory brane-world of Ref. \cite{ovrut}, which may
perhaps be relevant for the recently proposed Ekpyrotic universe scenario \cite{ekpy}, whereas
the RS model is recovered in the limit $q\to \infty$.

The plan of the paper is the following. Section II is devoted to the classical dynamics of the
model. After introducing the background solution, in Section II.b we derive the action for the 4D moduli fields.
These are massless at the classical level, due to a scaling symmetry of the 5D action which we
discuss in Section II.b. The hierarchy between the Planck scale and electroweak scale
is discussed in Section II.c, since in general there are some differences between
the present case and the RS case. In Section III we give a brief account of quantum effects by
considering the effective potential induced by a conformally invariant field.
This case is rather trivial, and it properly illustrates the Casimir interaction between the branes.
However, it misses the possibility
of interesting local terms induced at one loop. Section IV deals with the formalism for the
calculation of $V(\varphi_i)$ in more general cases, and contains the core of the discussion
on renormalization. We start, in Section IV.A, with a digression on the
specification of the measure of functional integration in a curved background. General
covariance alone is not sufficient to determine the measure in the present situation, due to
the presence of a nontrivial dilaton. Because of that,
we shall consider a one parameter family of possibilities, related with each other through
dilaton dependent conformal transformations.
Sections IV.B, C and D are devoted to dimensional regularization and zeta function regularization, which are shown
to give equivalent results. The explicit calculation
of the effective potential in some simple cases is given in Section V. The Heterotic M-theory model is briefly
discussed in Section VI. Possible consequences for the stabilization of the moduli are discussed
in Section VII, and our conclusions are summarized in Section VIII. Some technical issues
are left to the Appendices.

\section{The model}

The classical background we shall consider in this paper is a 5-dimensional spacetime with
a nontrivial background scalar field $\phi$, which we shall refer to as the ``dilaton''.
The fifth dimension is compactified on a $Z_2$ orbifold with two branes at the fixed
points of the $Z_2$ symmetry. The action for the background fields is given by
\begin{eqnarray}
    S_b &=& {-1\over \kappa_5}\int d^5 x\, \sqrt{g}
                \left( {\cal R}
        +\frac{1}{2}(\partial \phi)^{2}
        + \Lambda e^{c \, \phi}\right)\nonumber\\
    &-& \sigma_+ \int d^4x\sqrt{g_+}\,
    e^{c\,\phi/2} -\sigma_- \int_{}d^4x\sqrt{g_-}
    \,e^{c\,\phi/2},
    \label{general}
\end{eqnarray}
where $\cal R$ is the curvature scalar, $\kappa_5= 16\pi G_5$, where
$G_5$ is the 5-dimensional
gravitational coupling constant.
We have denoted the induced metrics on the positive and
negative tension branes by $g^+_{\mu\nu}$ and $g^-_{\mu\nu}$, respectively.
To find a solution of the equations of motion,
we make an ansatz where the 4-dimensional metric is flat,
\begin{equation}
    \label{metricform}
    ds^2=dy^2+a^2(y)\eta_{\mu\nu} dx^\mu dx^\nu,
\end{equation}
with a $x^{\mu}$-independent scalar field $\phi=\phi(y)$.
The positive and negative branes are placed at $y=y_+$ and
$y_-$, respectively.
Under these assumptions,
the equations of motion for $(a, \phi)$ in the bulk become
\begin{eqnarray}
&&\left({{\dot a}\over a}\right)^2 ={1\over 12}\;
\left({1\over 2}{\dot\phi}^2 - U(\phi)\right),\nonumber\\
&&\ddot\phi+4 \,{{\dot a}\over a}\,{\dot\phi}=U'(\phi),
\label{eom}
\end{eqnarray}
where $U(\phi)\equiv\Lambda e^{c \, \phi}$, a dot represents
differentiation with respect to $y$ and a prime
represents differentiation with respect to $\phi$.

As shown in \cite{youm},
there is a solution of Eqs.~(\ref{eom})
for any value of $c$ given by
\begin{eqnarray}
    \label{background}
    \phi&=&-{\sqrt{6{\kk}}}\, \ln (y/y_0), \cr
    a(y)&=&\left(y/y_0\right)^{{\kk}},
\label{solution}
\end{eqnarray}
where
\begin{equation}
{\kk}={2\over 3 c^2},\quad \quad
 y_0=\sqrt{3{\kk}(1-4{\kk})\over \Lambda}.
\label{defy0}
\end{equation}
(Constant rescalings of the warp factor are of course allowed,
but unless otherwise stated, we shall take the
convention that $a(y)=1$ at $y=y_0$.)
Assuming $y_-<y_+$, the boundary conditions which follow
from $Z_2$-symmetry imposed on both branes are given by
\begin{eqnarray}
   && \dot\phi_{\pm}=\mp {c\over4}\sigma_\pm
    e^{(c/2) \phi_\pm},\\
   &&
   6 \;{{\dot a}\over a}\Big|_\pm=\pm{1\over2} \kappa_5 \sigma_{\pm}
  e^{(c/2)
    \phi_\pm},
\end{eqnarray}
and they are satisfied if $\sigma_\pm$ are tuned to
\begin{equation}
\label{finetune}
\sigma_\pm=\pm {1\over \kappa_5}\sqrt{48{\kk}\Lambda\over1-4{\kk}}.
\end{equation}
In the absence of the branes, the spacetime (\ref{solution})
contains a singularity at $y=0$. Since we are considering
the range between $y_-$ and $y_+$, this singularity is of course inocuous.
Our spacetime consists of two copies of the slice
comprised between $y_-$ and $y_+$,
which are glued together at the branes.
Hence, the 5-th dimension is topologically
an $S^1/Z_2$ orbifold.

For $q=1/6$ this solution is precisely the M-theory heterotic
brane of Ref. \cite{ovrut}. On the other hand, the RS case, where
the bulk is AdS and there is no scalar field, can be obtained by
taking the limit $q\to \infty$ and $y_0\to \infty$ simultaneously,
while its ratio is kept fixed,
\begin{equation}
\ell=\lim_{q\to \infty} {y_0 \over q} = \sqrt{-12\over \Lambda}.
\label{ell}
\end{equation}
Defining $y \equiv y_0 + y^*$, we find that in the limit
the warp factor becomes an exponential
$$
\lim_{q\to \infty} a = e^{y^*/\ell},
$$
which corresponds to AdS space with curvature radius equal to
$\ell$.

\subsection{Moduli fields}

For fixed value of the coupling $c$, the solution given above
contains only two physically meaningful free parameters, which
are the locations of the branes $y_-$, and $y_+$.
This leads to the existence of the corresponding moduli,
which are massless scalar fields from the 4-dimensional point of view.
In addition to these moduli, the massless sector also contains the
graviton zero mode. To account for it, we generalize our metric
ansatz (\ref{metricform}) by promoting $\eta_{\mu\nu}$ to an
arbitrary four-dimensional metric:
\begin{equation}
ds^2= dy^2 + a^2(y) \tilde g_{\mu\nu}(x) dx^{\mu} dx^{\nu}.
\label{metric}
\end{equation}
It is easy to show that
${\cal R}= a^{-2} \tilde {\cal R} + {\cal R}^{(0)}$, where
${\cal R}^{(0)}$ is the background Ricci scalar in five dimensions
and $\tilde{\cal R}$ is the four-dimensional one.
For constant values of the metric and moduli, we have
a solution of the equations of motion whose action vanishes.
Hence, only the terms which depend on derivatives of the metric
or derivatives of the moduli will survive after the
five-dimensional integration. This fact can be used in order to
simplify the derivation of the effective action for the moduli,
since all terms without any derivatives can be dropped.

In the bulk, all terms will cancel except for the one which is
proportional to the four-dimensional Ricci scalar, $\tilde{\cal R}$.
Let us consider the contribution from the branes.
The metric induced on the branes is given by
$$
g_{\mu\nu}^{\pm} = a^2_{\pm}[\tilde g_{\mu\nu} + a^{-2}_{\pm}
\partial_\mu y_\pm \partial_\nu y_\pm].
$$
Here, and in what follows, the subindices $\pm$ mean that the quantity
is evaluated at the {\em perturbed} brane location.
The brane tension terms in the action contain the determinant
$$
\sqrt{- g_\pm} = a_\pm^4 \sqrt{-\tilde g} \left[1+{1\over
2 a^2_\pm}\ (\tilde\partial y_\pm)^2\right] + \cdots ,
$$
which induces kinetic terms for the moduli. Here, the tilde on the
kinetic term indicates that the derivatives are contracted
with the metric $\tilde g$.
In fact, the five dimensional Ricci tensor $\cal R$ contains second derivatives of the
metric and therefore it is singular on the brane, giving a finite
contribution to the action. To handle this contribution it is
simplest to introduce fiducial boundaries in the neighborhood of
the branes, were we add (back to back) pairs of Gibbons-Hawking boundary terms.
These have the form
\begin{equation}
{2 \over \kappa_5} \int d^4 x \sqrt{g_{\pm}}\
{\cal K_{\pm}},
\label{ghterm}
\end{equation}
where ${\cal K}_{\pm}$ is the trace
of the extrinsic curvature of the fiducial boundary near each one of the branes.
The action is separated into two parts. The first one is a ``bulk'' part,
consisting of an integral over two copies of an open set which excludes the branes,
$(y_- +\epsilon) < y < (y_+ - \epsilon)$, plus terms of the
form (\ref{ghterm}) at the boundaries $y=y_- + \epsilon$ and $y=y_+ -\epsilon$.
Then there is a ``brane'' part, which includes an integral of the action over the
infinitessimal open sets of thickness
$\epsilon$ around the branes, suplemented with terms of the form (\ref{ghterm}) at the boundaries
of these open sets (these ``brane'' boundary terms sit back to back with the ones
used in order to bound the ``bulk'', and have opposite sign relative to them, since the normal to
the fiducial boundary has opposite sign on each side of the boundary. Thus the total
effect of the boundary terms is to add zero to the action).
As is well known, through integration by parts
the boundary terms absorb the second
derivatives in the Einstein term ${\cal R}$. Hence, the singular contribution
from the gravity kinetic term on the brane disappears, and in the limit $\epsilon\to 0$,
the only contribution to the action from the
interval of width $2\epsilon$ around the branes (together with the added boundary terms),
is from the brane tension term itself, but not from the gravity kinetic term.
On the other hand, the bulk contribution has to be supplemented with the boundary terms
(\ref{ghterm}).

The boundary terms can be evaluated as follows.
Consider, for instance the hypersurface which is located at $y=y_+(x^\sigma)$.
In terms of the new coordinate $\hat y= y- y_+(x^\sigma)$,
the brane is at $\hat y=0$, and the metric can be written as
\begin{equation}
 ds^2=N^2 d\hat y^2+ g^+_{\mu\nu} (N^\mu d\hat y+dx^\mu)
          (N^\nu d\hat y+dx^\nu),
\end{equation}
where we have introduced the lapse function
$
 N^2=1- g^+_{\mu\nu}N^\mu N^\nu,
$
and the shift vector
$
 N^{\mu}=g^{+\mu\nu} y_{,\nu}.
$
Then, the trace of the extrinsic curvature is given by
\begin{eqnarray}
 \int dx^4 \sqrt{- g_+}\,{\cal K}_+  & = &
  \int dx^4 {\sqrt{- g_+}\over 2N}\left[
   g^{+\mu\nu}\partial_{\hat y}g^+_{\mu\nu}- 2N^\mu{}_{|\mu}\right]\cr
 & = &   \int dx^4 {\sqrt{g_+}\over 2N}\left[
   {\dot a\over a}\ g^{+\mu\nu} \tilde g_{\mu\nu}
    - {N^\mu \partial_\mu N^2\over N^2} \right]\cr
 & \approx & 4 \int \left({\dot a\over a}\right)_+ a_+^4\sqrt{-\tilde g}
     \left[1+{3\over 4a_+^2}
     (\tilde\partial y_+)^2 \right],
\end{eqnarray}
where the vertical line means a covariant differentiation
with respect to the induced metric $g^+_{\mu\nu}$.
We neglected the terms which are higher order in
in derivatives of the modulus $y_+$. As mentioned before, the subindices
$\pm$ mean that the quantities are evaluated at the {\em actual}
position of the brane, and the expression is in fact nonperturbative in the
positions $y_{\pm}$ themselves (although not in the derivatives).

Substituting the previous expressions into the action (\ref{general}),
with the addition of the extrinsic curvature terms,
and using the background equations of motion we find
\begin{eqnarray}
    S_b = \int d^4 x \sqrt{-\tilde g}\, \Biggl[
         &-&\left( 2 \int_{y_-}^{y_+} dy\,  a^2\right)
                             \frac{1}{\kappa_5} \tilde{\cal R}\cr
    & + &{1\over 2}\sigma_+ e^{(c/2) \phi_+}
        a^2_+\ (\tilde\partial y_+)^2
     + {1\over 2}\sigma_- e^{(c/2)\phi_-}
        a^2_-\ (\tilde\partial y_-)^2 \Biggr].
\label{moac}
\end{eqnarray}
This can be rewritten as
\begin{equation}
S_b={-1\over 16\pi G} \int d^4 x \sqrt{-\tilde g}
\left\{ (\varphi_+^2-\varphi_-^2)\ \tilde{\cal R} -
{6q\over q+1/2} \left[(\tilde\partial
\varphi_+)^2-(\tilde\partial\varphi
_-)^2\right]\right\}.
\label{moder}
\end{equation}
Here we have introduced
$$
\varphi_\pm \equiv \left({y_{\pm}\over y_0}\right)^{q+1/2},
$$
and the four dimensional Newton's constant $G$ given by
\begin{equation}
G = \left(q+{1\over 2}\right) {G_5\over y_0}.
\label{gn}
\end{equation}
The modulus corresponding to the positive tension brane has a
kinetic term with the "wrong" sign. However, this does not
necessarily signal an instability, because it is written in a
Brans-Dicke frame. One may go to the Einstein frame by a
conformal transformation. It is convenient to introduce the new moduli
$\varphi$ and
$\psi$ through \cite{ekpy}
$$
\varphi_+= \varphi \cosh \psi, \quad
\varphi_-= \varphi \sinh \psi,
$$
and to define the new metric
$$
\hat g_{\mu\nu} = \varphi^2 \tilde g_{\mu\nu}.
$$
It is then straightforward to show that $\varphi^2\ \sqrt{\tilde
g}\ \tilde{\cal R} = \sqrt{-\hat g}\ [\hat R + 6\
\varphi^{-2}\ (\hat\partial \varphi)^2]$. Substituting into
the background action (\ref{moder}), we have
\begin{equation}
S_b={-1\over 16\pi G} \int d^4 x \sqrt{-\hat g}
\left\{ \hat{\cal R} +\ {6\over 1+ 2q}\ {(\hat\partial
\varphi)^2\over \varphi^2} +\ {12 q \over
1+ 2q}\ (\hat\partial\psi)^2\ \right\}.
\label{einstein}
\end{equation}
Therefore, both moduli have positive kinetic terms in the Einstein
frame. At the classical level, the moduli are massless, but as we
shall see in the following Sections, a potential term of the form
\begin{equation}
\delta S= -\int d^4 x\ V(\varphi,\psi) \equiv -\int d^4x \sqrt{-\hat g}\ \hat
V(\varphi,\psi),
\end{equation}
is generated at one loop, which should be added to
(\ref{einstein}).

In the RS limit $q \to \infty$ [see Eq. (\ref{ell})] the kinetic
term for one of the moduli disappears. This is to be expected,
because the bulk is the maximally symmetric AdS
space. In this case only the relative position of the branes $y_+ - y_-$ is
physically meaningful and the other modulus can be gauged away (see
also \cite{barv} for a recent discussion of this case).

Matter does not couple universally to the 4D Einstein metric $\hat g_{\mu\nu}$. Rather, it
will couple to combinations of the metric and the moduli fields.
For completeness, in Appendix A we give an alternative form for
the moduli action in terms of the induced metrics on either brane,
to which brane matter fields are universally coupled.

\subsection{Scaling symmetry}

Since we are interested in the effective potential for the
moduli, it is perhaps pertinent to start by asking why these fields are
massless at the classical level. The reason is that under the global transformation
\begin{equation}
g_{ab} \to T^2 g_{ab},
\label{scaling1}
\end{equation}
\begin{equation}
\phi \to \phi- (2/c) \ln T,
\label{scaling2}
\end{equation}
the action (\ref{general}) scales by a constant factor
$$
S_b \to T^3 S_b.
$$
Here $g_{ab}$ is the metric appearing in the action (\ref{general}).
Acting on a solution with one brane, the transformation simply
moves the brane to a different location. Hence, all brane locations
are allowed, from which the masslessness of the moduli follows.
However, we should hasten to add that this is just a global
scaling symmetry which
need not survive quantum corrections.

It is interesting to observe that by means of a conformal transformation,
we may construct a new metric
$g^{(s)}_{ab}$ which is invariant under the scaling symmetry
\begin{equation}
g^{(s)}_{ab} = e^{c\phi} g_{ab}.\label{aki}
\end{equation}
In terms of this new metric the action takes the form
\begin{eqnarray}
    S_b &=& {-1\over \kappa_5}\int d^5 x\, \sqrt{g^{(s)}}
              e^{-3c\phi/2}  \left( {\cal R}^{(s)}
        +[(1/2)-3c^2](\partial^{(s)} \phi)^{2}
        + \Lambda \right)\nonumber\\
    &-& \sigma_+ \int d^4x\sqrt{g^{(s)}_+}\,
    e^{-3 c\,\phi/2} -\sigma_- \int_{}d^4x\sqrt{g^{(s)}_-}
    \,e^{-3 c\,\phi/2}.
    \label{general2}
\end{eqnarray}
Now, the symmetry is a mere shift in $\phi$.
Moreover, with our background solutions for $g_{ab}$ and
$\phi$, the metric $g^{(s)}_{ab}$ is just AdS, as can be easily shown from
(\ref{aki}) and (\ref{solution}).

For certain discrete values of $c$, the action (\ref{general2}) can be
obtained from dimensional reduction of $(5+n)$ dimensional pure gravity
with a cosmological term $\Lambda$, where the
additional $n$ dimensions are toroidal \cite{gpt3}. In this case,
the factor $e^{-3c\phi/2}$ is the overall scale of the internal
$n$-dimensional volume,
and the value of $c$ is given by $c^2=2/3q$, with
$$
q={n+3 \over n}.
$$
It is not surprising, then, that the metric $g^{(s)}_{ab}$ in
(\ref{general2}) corresponds to the ``external''
components of a $5+n$ dimensional anti de Sitter space,
since the starting point is in fact pure gravity in $(5+n)$ dimensions
with a negative cosmological term. The calculation of quantum corrections
in this higher dimensional space, and its relation with the calculation
of quantum corrections in the effective 5D theory which we consider in
this paper, will be reported in
a separate publication \cite{gpt3}.

\subsection{The hierarchy}

As mentioned in the introduction, one of the motivations for
studying brane-world scenarios has been the search for a geometric
origin of the hierarchy between the effective Planck scale $m_p$ and the
electroweak scale. In the 5D description, all matter fields are assumed
to have masses which are close to the cut-off scale of the theory $M\equiv G_5^{-1/3}$.
And yet, with the help of an exponential warp factor (as in the RS model) it is easy to
generate a hierarchy of the order of $m_p/m \sim 10^{16}$.
Here $m$ is the effective mass of fields which live on the
negative tension brane, as ``seen'' in the effective four dimensional
description \cite{RS1}. In this subsection we shall review this mechanism,
including the case of a warp factor with
arbitrary power $q$, since there are some minor differences with the RS case.

The effective four-dimensional Planck mass is given by
\begin{equation}
m_p^{2} = {2\over 1+2q} {M^{3} y_+}
\left[1-\left({y_-\over y_+}\right)^{2q+1}\right],
\label{massp}
\end{equation}
where $M$ is the 5-dimensional Planck mass, as can be seen from
Eqs. (\ref{moder})
and (\ref{gn})], where, without loss of generality, we have taken
$y_+=y_0$. Here, and for the rest of this Section, we shall follow standard
practice and refer
all physical quantities to the measurements of clocks and rods located on the
positive tension brane.

Let us now consider the mass scales of fields which live on the branes.
We expect these fields to couple not only to
the metric, but also to the background dilaton $\phi$. There
are many possible forms for this coupling, but it seems reasonable to
restrict attention to those which respect the scaling symmetry
(\ref{scaling1}-\ref{scaling2}). For a free scalar field $\Psi$
which lives on the negative tension brane, and whose
mass parameter is comparable to the cut-off scale, the
action takes the form
\begin{equation}
S_\Psi = -{1\over 2}\int \sqrt{g^{(s)}_-}
F^2(\phi) [g^{(s)\mu\nu}_- \partial_\mu \Psi
\partial_\nu \Psi + \alpha M^2 \Psi^2].
\label{similar}
\end{equation}
Here we have introduced a fudge factor $\alpha$ to allow for an
intrinsic mass which is slightly lower than the cut-off scale.
The function $F(\phi)$ can be reabsorbed in a redefinition of $\Psi$,
and thus the relevant warp which determines the hierarchy between
mass scales on the positive and in the negative tension branes is
the one corresponding to the metric $g^{(s)}$.
Our field will be perceived from the point of view of the 4D effective
theory as having a mass squared of order
\begin{equation}
m^2\sim \alpha M^2 \left({y_-\over y_+}\right)^{2q-2}.
\label{massn}
\end{equation}
Notice that there are two different factors which determine the hierarchy
between $m$ and $m_p$.
The first one is the warp factor $(y_-/y_+)^{q-1}$ appearing in
Eq. (\ref{massn}), which ``redshifts'' the mass scales of particles on
the negative tension brane (except for $q<1$, in which case the
particles on the negative tension brane appear to be heavier than
those on the positive tension brane). This generates the hierarchy in the
RS model. The second one is the possibly large volume of the
internal space, which enhances the effective Planck scale with
respect to the cut-off scale [see
Eq. (\ref{massp})]. This generates the hierarchy in
the ADD model \cite{gia} with large extra dimensions.
Considering both effects, the hierarchy $h$ is given by
$$
h^2 = {m^2 \over m_p^2} \sim \alpha\ {1+2q \over 2}{1\over M y_+} \left
({y_-\over y_+}\right)^{2q-2}.
$$
It is known that
without a warp factor, it is not possible to generate the desired
hierarchy from a single extra dimension, since its size would have
to be astronomical.
An interesting question is what is the minimum
value of the exponent $q$ which would be sufficient in order to generate a
ratio $m/m_p \sim 10^{-16}$. The best we can do is to take the curvature
scale $(y_+/q)$
slightly below the millimeter scale,
\begin{equation}
(y_+/q)\lesssim m_p (TeV)^{-2} \sim mm.
\label{upper}
\end{equation}
in order to pass the short distance tests on deviations from
Newton's law. On the other hand, we also need
\begin{equation}
(y_-/q) \gtrsim M^{-1},
\label{lower}
\end{equation}
since for smaller values of $y_-$ the curvature becomes comparable
to the cutoff scale $M$ and the theory cannot be trusted
\footnote{One should also bear in mind that the scale $M$ might
itself be a ``derived'' quantity,
as it happens for instance with the Planck mass in theories with additional
large extra dimensions. In this case, the true
cut-off scale may well be below $M$. Hence, the lower bound
(\ref{lower}) on $y_-$ should just be considered a necessary condition
for the low energy  description not to break down.}. Substituting
in (\ref{massp}) we have $M^3 \gtrsim m_p (TeV)^2$ and
$(y_-/y_+)\gtrsim (m_p/TeV)^{-4/3}$, which leads to
\begin{equation}
10^{-32}\sim {m^2\over m_p^2} \gtrsim \alpha
\left({TeV\over m_p}\right)^{4(2q-1)\over 3}.
\label{limq}
\end{equation}
Hence, a warp factor with exponent $q\geq 5/4$
may account for the observed hierarchy with a single extra
dimension, but it appears that this cannot be done for lower values
of $q$. \footnote{Except, of course, by giving up the assumption that
the Lagrangian of matter on the branes should scale in the same way
as the rest of the classical
action [see the discussion around Eq. (\ref{similar})]. If we allow
any coupling of $\phi$ to the mass term for $\Phi$, then any hierarchy
can be easily generated for any value of $q$.}
In particular, the Heterotic M-theory model, with $q=1/6$,
does not seem to allow for such possibility.

\section{Effective potential induced by bulk conformal fields}
\label{sec:confc}

Before embarking on a detailed discussion of renormalization, we shall consider in this
Section the case of a conformal bulk field $\chi$.
This case is rather easy to handle, and it is useful in
illustrating the non-local contribution of the vacuum energy to
the masses of the moduli. Unfortunately, it misses possibly interesting
effects which may arise due to renormalization of local operators
on the brane, whose discussion we defer to the next sections.


Following the discussion given in \S~III of \cite{gpt},
we define the conformal coordinates by
\begin{equation}
 z\equiv \left|\int {dy\over a(y)}\right|={y_0\over |1-{\kk}|}\left({y\over
 y_0}\right)^{1-{\kk}},
\end{equation}
and we rewrite the metric as
\begin{equation}
    \label{metricconf}
    ds^2=a^2(z)\big(dz^2 + \eta_{\mu\nu} dx^\mu dx^\nu
      \big),
    \quad\qquad a(z)=(z/z_0)^\beta,
\label{conformalmetric}
\end{equation}
where
\begin{equation}
\beta={{\kk}\over1-{\kk}}, \quad\qquad z_0={y_0\over |1-{\kk}|}.
\label{defz0}
\end{equation}
Here we should mention that the direction of increasing $z$
does not coincide with the direction of increasing $y$
when $q>1$.

The Casimir energy density $\rho$ of the conformally coupled scalar field
in this spacetime is related to its counterpart
in a flat spacetime with a compactified extra dimension
($a=1$ in the above metric) by a conformal transformation.
The relation is $\rho=a^{-5}\rho_0$, where $\rho_0$ is the
flat space value:
\begin{equation}
    \label{rrr}
    \rho_0 = {V_0 \over 2 |z_+ - z_-|} = \mp {A \over 2
    |z_+-z_-|^5};
    \qquad\qquad\quad  A\equiv{\pi^2 \over 32} \zeta'_R(-4)
    \approx 2.46 \cdot 10^{-3}.
\end{equation}
Here $z_+$ and $z_-$ are the position of the positive and
the negative tension branes in the conformal coordinate $z$.
The double signs in (\ref{rrr}) refer to bosons or fermions respectively.
Then, we find that the contribution of a conformally coupled
scalar field to the effective potential per unit co-moving volume
is given by
\begin{equation}
    V(z_+,z_-) = \sqrt{-\hat g}\ \hat V= 2\int a^5(z) \rho\, dz=
\mp {A \over |z_+ - z_-|^4}.
\label{ve1}
\end{equation}
Here, $\hat V$ is the effective potential per
unit "physical" volume (i.e. the volume as measured with the
Einstein metric $\hat g$), to be inserted in (\ref{einstein}).
In terms of the moduli fields $\varphi$ and $\psi$ we have
\begin{equation}
\hat V(\varphi,\psi)=\mp\ B\ \Lambda^2 \varphi^{-12\gamma}
\left[(\cosh\psi)^{3\gamma-1} -(\sinh\psi)^{3\gamma-1}\right]^{-4},
\label{hattie}
\end{equation}
where we have defined
$$
B= { A (1-q)^4
\over 9 q^2(1-4q)^2}, \quad \quad \gamma={1\over 1+2q},
$$
and we have used $\sqrt{\hat g}= \varphi^4$ for the background
solution.
It should be mentioned that local terms induced by quantum corrections
may be added to (\ref{hattie}), both due to fields which
live on the branes as well as from non-conformal bulk fields. A
discussion of these terms is deferred to the next section.


In the RS case $\gamma\to 0$ the $\varphi$ dependence in (\ref{hattie})
disappears, and we recover the results of \cite{gpt}. We have
$$
\hat V_{RS} = \pm {A\over 16 \ell^4} e^{4\psi}\sinh^4  2\psi.
$$
For $\psi\ll 1$,
the field $\psi$ corresponds to the hierarchy
between scales on both branes:
$$
\tanh\psi\approx \psi = {a_-\over a_+} \sim {TeV \over m_{Pl}} \ll
1,
$$
where $m_{Pl}\sim 10^{19} GeV$ is the Planck mass.
In the second equation we assume that the warp is responsible for
the hierarchy between Planck and electroweak scales. In this case we have
$$
\hat V_{RS}(\psi) \approx {A\over \ell^4}\ \psi^4 [ 1+  4 \psi + ...] \sim A
\ (TeV)^4.
$$
Here we have assumed that the AdS radius is not far below the
Planck scale, in which case the contribution to the effective potential
is of electroweak order. This huge contribution must somehow be cancelled
by some other term in order to have an acceptable four dimensional
cosmological constant. Also, the slope of the potential has to be
cancelled at the place where the radion sits. These two conditions
can be imposed if we allow for a finite
renormalization of the brane tensions, which contribute
proportionally to $a_+^4 \propto \cosh^4\psi\approx 1$ and $a_-^4\propto \sinh^4\psi \approx \psi^4$
for the positive and negative tension branes respectively.
With these additions we have
$$
\hat V_{RS}(\psi) = A \ell^{-4}\left[c_1 + c_2 \psi^4 + 4 \psi^5 + ...\right],
$$
where $c_1$ and $c_2$ are undetermined constants which can only be
fixed by experiment. The "renormalization" condition
$\partial_\psi V(\psi) = 0$ at the observed value of the radion
$\psi=\psi_{obs}\sim TeV/m_{Pl}$ forces $c_2 \sim \psi_{obs}$.
Hence the mass of the radion induced by this quantum correction is given by
\cite{gpt}
\begin{equation}
m^2_\psi \approx\ A \ {\psi_{obs}^3\over \ell^{4} m_{Pl}^2} \sim A\
\psi_{obs} (TeV)^2.
\label{massco}
\end{equation}
Again, in the last equality, we have taken $\ell$ to be similar to
the Planck scale. The radion mass is much lower than the electroweak
scale and hence it
is not enough for a phenomenologically viable stabilization
of the radion. The reason for such as small mass is that
$V(\psi)$ is polynomial for small $\psi$, with the first
power being $\psi^4$. The
situation would be different if the potential took the form
$\psi^4\log(\psi)$ as in the Coleman-Weinberg case. Then,
the induced mass $m^2_\psi$ would be of the same order as
the induced potential. Remarkably, the absence of logarithmic
terms seems to extend to the contribution from massive fields too
\cite{gr}, and so in the RS case it appears that the radion cannot
acquire a phenomenologically interesting mass solely from
one loop quantum effects.

For the more general case of a power
law warp factor, the possibility of an efficient stabilization by
quantum effects will depend on the nature
of the local operators which are induced by quantum corrections.
We now turn to a discussion of this subject.

\section{Effective potential in the general case}
\label{sec:nonmin}

In this Section we set up the framework for computing the
contribution to the 1-loop effective potential
from a scalar field $\chi$ propagating in the bulk
with a generic mass term, which may include couplings to the
curvature of spacetime as well as couplings to the background
dilaton $\phi$. The effective potential for the moduli $y_{\pm}$ will be defined as usual
in terms  of a Gaussian path integral around the background solution. Before presenting
the actual calculation, however, a digression on the choice of
the measure of integration will be useful.

\subsection{Specification of the functional measure}

A quantum field theory
is defined not just by the classical Lagrangian, but it is also
necessary to specify the measure of functional integration.
The latter is usually prescribed by demanding certain symmetries
or invariances. For instance, for scalar fields in curved space,
invariance under diffeomorphisms is an obvious requirement. If gravity
were the only background field, then this requirement would suffice
to uniquely define Gaussian integration around that background.
On the other hand, if there are fields other than gravity with a
nontrivial profile (such as our dilaton $\phi$), then there is
a wide class of possible choices, related to each other by dilaton dependent conformal
transformations. All choices within this class are equally good from
the point of view of diffeomorphism invariance. Strictly speaking, however,
they are inequivalent due to the well known conformal anomaly.

To be definite, let us concentrate in the simple case
of a bulk scalar field $\chi$ with canonical kinetic term.
The (Euclidean) action for this field is given by
\begin{equation}
    S[\chi]={1\over 2}\int d^{D}x
     \sqrt{g}\,\chi P \chi \,,
\label{Schi}
\end{equation}
where we have introduced the covariant operator
$$
P=-(\Box_g+ E).
$$
Here $\Box_{g}$ is the d'Alembertian operator
associated with the metric
$g_{ab}$, and $E=E[g_{ab},\phi]$ is a generic ``mass'' term.
Typically, this takes the form $E=-m^2-\xi {\cal R}_g$, where
$m$ is a constant mass, ${\cal R}_g$ is the curvature
scalar and $\xi$ is an arbitrary coupling.
Throughout this Section we shall leave $E$ unspecified.

A volume measure ${\cal D}\chi$ in field space ${\cal F}$
can be found from a metric $G_{xy}$ on ${\cal F}$,
through the relation:
\begin{equation}
{\cal D} \chi = \sqrt{G}\  \prod_x d\chi^x.
\label{defmeasure}
\end{equation}
Here, the spacetime coordinates $x$ and $y$ are considered as
continuous labels for the coordinates
$\chi^x\equiv \chi(x)$ of the infinite dimensional space ${\cal F}$, and $G$ is the
determinant of $G_{xy}$.
To specify $G_{xy}$, we note that a natural definition of a scalar
product in the space of field variations
$\delta\chi$ can be given in terms
of the spacetime measure $d\mu(x)$, through the relation
$$
<\delta\chi_1,\delta\chi_2>_\mu
\equiv
\int \int d\mu(x) d\mu(y)\ G_{xy}\ \delta\chi^x_1\ \delta\chi^y_2
\equiv
\int d\mu(x) \delta\chi_1(x)\delta\chi_2(x).
$$
We denote field variations
by $\delta\chi$ just to emphasize that we are refering to elements of the
tangent space.
More precisely, $\delta\chi= \delta\chi^x e_x$, where
$e_x=\partial/\partial\chi^x$ is the coordinate
basis of the tangent space at the point $p$ which corresponds to the
background solution.
In a Riemannian spacetime,
the invariant measure is given by
\begin{equation}
d\mu(x)=\sqrt{g(x)}d^D x,
\label{stmeasure}
\end{equation}
where $g$ is the determinant of
$g_{ab}$, and $D$ is the dimension. The
implicit definition of $G_{xy}$ given above is just the identity
$\delta_\mu(x,y)$ with
respect to $d\mu$ integration,
\begin{equation}
G_{xy}=\delta_{\mu}(x,y) = {\delta^{(n)}(x-y)\over \sqrt{g(x)}}.
\label{fieldmetric}
\end{equation}
It is convenient to express the field variations in an orthonormal basis
${\chi_n}$, with $<\chi_n,\chi_m>=\delta_{nm}$, so that
$\delta\chi(x)=\sum_n c^n \chi_n(x)$. In this basis, the
components of the field variation are $c^n$, and the metric is just
the usual delta function (the continuous or
the discrete delta function depending
on whether the normalization of $\chi_n$ is continuous or discrete):
$$
G_{nm}=\delta_{nm}.
$$
Substituting in (\ref{defmeasure}), we have
\begin{equation}
{\cal D}\chi = \prod_n dc^n.
\label{orthomeasure}
\end{equation}

It should be clear from the previous discussion that the definition of
${\cal D}\chi$ is associated with a natural definition
of $d\mu(x)$. However, in the problem
under consideration in this paper, the choice of $d\mu$ is not
unique. In our case, there is a nontrivial
dilaton field $\phi$, and we can consider a whole class
of spacetime measures of the form
$$
d\mu_\theta(x) = \sqrt{g_\theta}\ d^D x =
\Omega^D_\theta(\phi)\sqrt{g}\ d^D x,
$$
which correspond to conformally related metrics
$$
g^\theta_{ab}=\Omega_\theta^2 g_{ab},
$$
for an arbitrary function $\Omega_\theta(\phi)$ (we shall use some
specific choices of this function in the next subsection). In the
presence of a dilaton, the coupling to gravity is not universal and
it is not clear which one of these metrics should be considered more physical.
To proceed, it is convenient to define the operator $P_{\theta}$ associated with the
metric $g^{\theta}_{ab}$ by
\begin{equation}
\Omega_\theta^{(D-2)/2}{P_\theta}
\Omega_\theta^{(2-D)/2}=
\Omega_{\theta}^{-2} P,
\label{allsorts}
\end{equation}
where $P$ was introduced after Eq. (\ref{Schi}).
This operator can be written in covariant form
as
$$
P_{\theta}= - (\Box_{\theta}+ E_{\theta}),
$$
where
$$
{E_{\theta}}=\left({D-2\over 2}\right)\ \Box_\theta\ln\Omega_\theta-
\left({D-2\over 2}\right)^2 g^{ab}_\theta\ \partial_a \ln\Omega_\theta\
\partial_b \ln\Omega_\theta
  + \Omega_\theta^{-2} E,
$$
and $\Box_\theta$ is the covariant d'Alembertian corresponding to
$g^\theta_{ab}$.
Introducing $\chi_\theta \equiv \Omega_\theta^{(2-D)/2} \chi$, the action for the
scalar field can be expressed as
\begin{equation}
S[\chi]={1\over 2}\int d^{D}x\sqrt{g_\theta}\ \chi_\theta\
    P_\theta\ \chi_\theta .
\label{Shatchi}
\end{equation}
In terms of $g^{\theta}_{ab}$ the field $\chi_\theta$ has a perfectly canonical and covariant
kinetic term.

Thus, the same arguments which lead to
(\ref{fieldmetric}) can now be used in order to find the natural line element
in field space associated with the spacetime measure $d\mu_{\theta}(x)$:
$$
d{\cal S}^2_\theta = \int \int d\mu_\theta(x)d\mu_\theta(y)
G^\theta_{xy}\  d\chi_\theta^x\ d\chi_\theta^y\
+ ...\ =
\int d^Dx{\sqrt{g_\theta(x)}}\ \left(d\chi_\theta(x)\right)^2+ ...
$$
Here, the ellipsis denote the omitted terms which
correspond to variations of other fields in the theory (in particular,
these include the variations of the gravitational field and the dilaton).
Let us compare this line element with the one considered above
$$
d{\cal S}^2=\int \int d\mu(x)d\mu(y) G_{xy}\ d\chi^x\ d\chi^y\ + ...=
\int d^Dx{\sqrt{g(x)}}\ \left(d\chi(x)\right)^2 + ...
$$
For field variations where $\chi$ changes
but the rest of the fields (metric, dilaton, etc.) are constant, we have
\begin{equation}
d{\cal S}^2_\theta = \int d^Dx\ \Omega_\theta^{2}{\sqrt{g(x)}}\ \left(d\chi(x)\right)^2 \neq d{\cal S}^2 \quad \quad
(g, \phi, ...=const. ; d\chi^x\neq 0),
\label{theirdif}
\end{equation}
and therefore $d{\cal S}^2_\theta \neq d{\cal S}^2$ in general. Of course,
the corresponding measures of integration will also be different.
In the basis $\{\chi_{\theta n}\}$ which is orthonormal with respect
$d\mu_\theta$, the field variation can be expanded as $\delta\chi_\theta(x)=
\sum_n c^n_\theta \chi_{\theta n}$, and the new measure takes the form
\begin{equation}
\left({\cal D}\chi\right)_\theta = \prod_n dc^n_\theta.
\label{orthomeasuretheta}
\end{equation}
Using $\chi_{\theta m} = \Omega_\theta^{-D/2} \chi_m$, it is straigtforward
to show that $c^m = M^m_n c^n_\theta$, where
$M^m_n = <\chi_m, \Omega_\theta^{-1} \chi_n>_\mu\equiv (\Omega_\theta^{-1})^m_n$.
Hence the two measures (\ref{orthomeasure}) and (\ref{orthomeasuretheta}) are
related by
\begin{equation}
{\cal D}\chi = J_\theta\ ({\cal D}\chi)_\theta,
\label{change}
\end{equation}
where the Jacobian is formally given by
\begin{equation}
J_\theta= \det(\Omega_\theta^{-1})=
\exp\left[{-{\rm Tr} \ln \Omega_\theta}\right].
\label{functionaljacobian}
\end{equation}
In the last step we have used the formal definition
of the $L_2$ trace:
\footnote{The definition of the trace is robust, in the sense
that it is independent on  the metric one uses in order to define the
orthonormal basis, as
long as the corresponding measures are in the same $L_2$ class. This
will be the case, for instance, if the metrics are related by a conformal
factor which is bounded above and below on the manifold.} :
$$
{\rm Tr}[{\cal O}]=
\sum_m\int d^D x\ { g}^{1/2}\  \chi_m ({\cal O}\chi_m)=
\sum_m\int d^D x\ {g_\theta}^{1/2}\  \chi_{\theta m} ({\cal O}\chi_{\theta m}).
$$
The trace is well defined if the diagonal matrix
elements of the operator ${\cal O}$ decay sufficiently fast
at large momenta. Unfortunately, the diagonal matrix elements of
$\ln \Omega_\theta$ do not decay at all at large $m$, and so the trace
is ill defined unless we introduce a regulator. We will address this
question below, where we will explicitly define what we mean by $J_\theta$.

Perhaps we should add, for the sake of clarity, that the difference
between the line elements
$d{\cal S}^2$ and $d{\cal S}^2_\theta$, and consequently the difference
between the associated measures, {\em is not due to field
redefinitions}. Both objects are different, but since they are defined
geometrically, they are both invariant under field redefinitions (in the same
sense that any line element is invariant under coordinate transformations).
Rather, the relation (\ref{functionaljacobian}) expresses the well known conformal
anomaly. The measure is not invariant under conformal transformations because these
do not correspond to a change of coordinates in field space ${\cal F}$. They correspond to a change of the
spacetime metric and consequently to a change of the metric on ${\cal F}$. This sort
of ambiguity does not arise when we consider scalar fields in flat space.
Consider an action with a general kinetic term of the form:
$$
S=\int d^D x\ G_{AB}(\phi^C) \eta^{ab} \partial_a \phi^A\partial_b\phi^B + ...
$$
A natural line element in field space can be obtained by
``stripping off'' the flat metric
$\eta^{ab}$ and replacing the partial derivatives with differentials of
the fields:
$$
d{\cal S}^2 = \int d^D x\ G_{AB}(\phi^C) d\phi^A d\phi^B.
$$
This procedure cannot be transported into a curved space, because
in erasing the factor $g^{ab}$ from the kinetic term, it makes a difference
what exactly we have chosen to call the metric of spacetime: $g_{ab}$ or
$g^\theta_{ab}$ (this is, by the way, the reason why the factor of $\Omega_\theta^2$
appears in (\ref{theirdif})). Usually, flat space definitions can be generalized
to curved space through the principle of general covariance: objects should be defined
geometrically, and they should reduce to their flat space definition when the spacetime
metric is flat. The question is, however, which object should be considered to
play the role of spacetime metric, so that we know when to call it flat.
Physically, too, one should expect that a preferred spacetime metric should play a role
in regularizing and renormalizing the theory. Suppose that we attempt to regularize
with a physical cut-off, so that all degrees of freedom beyond
a certain scale are ignored. In our background (which is conformally
flat) a constant physical cut-off scale corresponds to a different co-moving
scales at different places. The relation between physical and co-moving
scale is of course given by the metric, and therefore it can make a difference
which one we use.

Since we have a classical scaling symmetry in the gravity and dilaton sector,
one could argue that $g^{(s)}_{ab}$, which is invariant under scaling (see Section II.b),
is the preferred physical metric. \footnote{Note, in particular, that the overall
scaling factor of the action (\ref{general}) under (\ref{scaling1})
and (\ref{scaling2}) depends on the spacetime dimension, and hence
the symmetry itself is different when we change the dimension.
By contrast, the scaling of (\ref{general2}) remains the same
in any dimension.}
However, even in this case the
divergent part of the effective potential will not respect the
scaling symmetry, and consequently we need to introduce counterterms with the
"wrong" scaling behaviour. Hence, in what
follows, we shall take the conservative attitude that the
measure is determined in the context of a more
fundamental theory (from which our 5-D effective action is derived),
and we shall formally consider on equal footing all choices associated with
metrics in the conformal class of $g_{ab}$, including of
course $g^{(s)}_{ab}$. As we shall see, the difference between these choices
amounts to the addition of local terms in the effective potential.

The contribution of the field $\chi$
to the renormalized effective potential $V_\theta$ per unit co-moving volume
parallel to the branes is given by:
\begin{equation}
\exp\left[-{\cal A} (V_\theta + V^{div}_\theta)\right] \equiv
\int ({\cal D}\chi)_\theta\ e^{-S[\chi]} = ({\rm det} P_\theta)^{-1/2},
\label{veffdef}
\end{equation}
where ${\cal A}$ is the co-moving volume under consideration and we have used (\ref{Shatchi})
and the measure (\ref{orthomeasuretheta}) to express the gaussian integral as a determinant.
The term $V^{div}_\theta$ is a
local counterterm which, in dimensional regularization,
needs to be subtracted from the regularized
effective potential (its explicit form will be given in the
coming sections). In zeta function regularization, the left
hand side is already finite, and $V^{div}_\theta$ is unnecessary (it
corresponds to a finite renormalization of couplings).
Eq. (\ref{veffdef}) can be written as
\begin{equation}
V_\theta \equiv {1\over 2{\cal A}} \,{\ln}(\det P_\theta) - V_\theta^{div}.
\label{newveffdef}
\end{equation}
Eq. (\ref{change}) suggests the notation
\begin{equation}
V_\theta = {1\over 2{\cal A}}\ln(\det P)
- V_\theta^{div} +
{1\over{\cal A}}\ln J_\theta .  \label{48'}
\end{equation}
The reader should be aware, however, that
the definition of the Jacobian in Eq. (\ref{functionaljacobian}) is only formal
because the trace in the r.h.s. of this equation is ill defined.
For that reason, it is not clear that Eq. (\ref{48'}) would hold with the
definition (\ref{functionaljacobian}), after substituting
determinants by traces and applying any kind of regularization to the formally
divergent traces. To avoid misinterpretations, in the discussions that follow
we shall take $J_\theta$ to be defined by Eq. (\ref{48'}), that is
\begin{equation}
  {\ln} J_\theta
      \equiv {1\over 2}\left[{\ln}(\det P_\theta)
        -{\ln}(\det P)
        \right],
\label{TrlnOmega}
\end{equation}
where the expression in the right hand side is to be calculated in some regularization
scheme.

The way the $\theta$ dependence of $V_\theta$ arises is very different in
different regularization schemes. In Eq. (\ref{48'}),
the determinant of $P$ is independent of $\theta$ (we recall that
this operator corresponds to the choice $\Omega_\theta =1$).
In dimensional regularization, $\ln J_\theta$ vanishes,
but the divergent term  $V_\theta^{div}$ which is subtracted from
$\ln(\det P)$ depends on the choice of physical metric $g^{\theta}_{ab}$.
On the other hand, in zeta function regularization, $\ln(\det P)$
is finite and
$V^{div}$ does not play a role (in any case, any finite renormalization does
not introduce a dependence in $\theta$). Rather, in this case, the dependence
on $\theta$ comes from $\ln J_\theta$,
which does not vanish in this regularization scheme. In both cases, the
$\theta$ dependence of $V_\theta$ is the same.

As we shall see, this dependence can
be cast in the form of local operators on the branes,
and therefore the ambiguity in the
choice of the integration measure can also be understood
as modification of the classical action. It should be noted, however,
that the local operators which result from a shift in $\theta$ have different
form than the terms arising from the usual shift in
the renormalization constant $\mu$ which inevitably crops up in the regularized traces.
In the cases we shall consider, the latter will take the
form $K^4(y_\pm)$, where
$K$ denotes terms which behave like the extrinsic curvature of the branes at the positions $y_\pm$.
On the other hand, the $\theta$-dependent terms behave as $K^4(y_\pm)
\phi(y_\pm)$. Since $K(y)$ behaves like the inverse
of $y$ whereas $\phi(y)$ behaves logarithmically with $y$, these terms
will give rise to Coleman-Weinberg
type potentials for the moduli.

\subsection{Conformal transformations and the KK spectrum}

The direct evaluation of the determinant of $P_\theta$ appearing in
Eq. (\ref{veffdef}) turns out to be rather impractical,
due to the complicated form of the implicit equation which defines
its eigenvalues. For actual calculations it is convenient to
work with a conformally related operator $P_0$ whose eigenvalues
will be related to the KK masses.

Following~\cite{gpt},
we introduce a one-parameter family of
metrics which interpolate between a fictitious flat spacetime
and any of the metrics in the conformal class of the Einstein metric:
\begin{equation}
g^{\theta}_{ab}=\Omega_{\theta}^2(\phi)\, g_{ab},
\label{confT}
\end{equation}
where $\theta$ parametrizes a path in the space of conformal
factors.
For definiteness we shall restrict attention to
conformal factors $\Omega_\theta(\phi)$
which have an exponential dependence
on the dilaton:
\begin{equation}
    \Omega_{\theta}(z)=e^{(1-\theta) \phi/3c}=\Big({z\over z_0}\Big)^{\beta\,(\theta-1)}.
\label{conformali}
\end{equation}
With this choice, $\theta=0$ represents flat space and $\theta=1$
corresponds to the Einstein frame metric (\ref{conformalmetric}).
For $\theta=-1/\beta$, the metric $g^{\theta}_{ab}$ coincides
with the metric $g^{(s)}_{ab}$ introduced in Subsection II.B, which
is invariant under the scaling transformation (as mentioned before,
this metric corresponds to a five dimensional AdS space, with
curvature radius given by $z_0$).

The operator $P_0\equiv P_{\theta =0}$ is the wave
operator for the KK modes which one would use in a four-dimensional description.
The Lorentzian equation of motion $P \chi=0$ can be written as
$$
P_0 \chi_0=0,
$$
where
\begin{equation}
P_0 = -\Box_{D-1} + \hat M^2 (z).
\label{pobox}
\end{equation}
Here $\Box_{D-1}$ is the flat space d'Alembertian
along the branes, and
$$
\hat M^2 \equiv -\partial_z^2 - E_0,
$$
is the Schr{\"o}dinger operator whose eigenvalues are commonly
referred to as the KK masses $m_n$:
\begin{equation}
\hat M^2(z) \chi_{0,n}(z) = m^2_n \chi_{0,n}(z).
\label{massess}
\end{equation}
The interesting feature of (\ref{pobox}) is that it separates into
a four-dimensional part and a $z$ dependent part.
A mode of the form $\chi_0= e^{ik_{\mu}x^{\mu}} \chi_{0,n}$ will
solve the equation of motion (\ref{pobox}) provided that the
dispersion relation
$$
k_{\mu}k^{\mu} + m_n^2 =0,
$$
is satisfied, and hence modes labeled by $n$ behave as four-dimensional
massive particles. Technically, the advantage of working with $P_0$ is that
its (Euclidean) eigenvalues
$\lambda_{n,k}=k_{\mu}k^{\mu} + m_n^2$ separate as a sum of a
four-dimensional part plus the eigenvalue of the Schr{\"o}dinger
problem in the fifth direction.

In the following subsections, we shall discuss how
${\rm det} P_0$ is related to the determinant of our interest,
${\rm det} P$, or more generally to ${\rm det} P_\theta$. For completeness,
and in order to illustrate practical methods for calculating the effective
potential, we shall consider dimensional regularization and zeta function
regularization. Both methods will lead to identical results.

\subsection{Dimensional regularization}

A naive reduction to flat four-dimensional space
suggests that the effective potential can be obtained as a
sum over the KK tower:
\begin{equation}
V^{D} \equiv \mu^{\epsilon}\sum_n{1\over 2}\int {d^{D-1}
      k\over
      (2\pi)^{D-1}}\log\left({k^2 + m_n^2(\varphi_i,D)
      \over\mu^2}\right).
\label{4b}
\end{equation}
Here $D=4+1-\epsilon$ is the dimension of spacetime, and we have
added ($-\epsilon$) dimensions parallel to the brane. The renormalized
effective potential should then be given by an expression of the form
\begin{equation}
V(\varphi_i) = V^D-V^{div},
\label{4a}
\end{equation}
and the question is what to use for the divergent
subtraction $V^{div}$.
Since Eq. (\ref{4b}) is similar to an ordinary effective potential
in 4-dimensional flat
space\footnote{
It should be mentioned also that each KK contribution in
Eq.(\ref{4b})
is not just like a flat space contribution, because in
warped compactifications the KK
masses $m_n(\varphi,D)$ depend on the number of external dimensions
parallel to the brane.},
one might imagine that $V$ can be obtained from $V^D$
just by dropping
the pole term, proportional to $1/\epsilon$; but this is not true
for warped compactifications
$$
V(\varphi_i) \neq V^{D}- {(\rm pole\ term)}.
$$
The point is that the theory is five dimensional and the spacetime is curved,
and this fact must be taken into account in the process of renormalization.

Rather than proceeding heuristically from (\ref{4a}), we must take
the definition of the effective potential Eq. (\ref{newveffdef}) as our
starting point,
where it is understood that the formally divergent trace must be regularized
and renormalized. In order to identify the divergent quantity to
be subtracted, we shall use standard heat kernel expansion techniques.
Let us introduce the dimensionally regularized expressions \cite{odintsov}
\begin{equation}
    V^{D}_{\theta}\equiv {\mu^{\epsilon}\over 2{\cal A}} {\rm Tr}\,{\ln}
    \left({P_{\theta}(D)\over \mu^2}\right)
    =-{\mu^{\epsilon}\over 2 {\cal A}}\lim_{s\to 0}\partial_{s}
    \zeta_{\theta}(s,D),
    \label{vthetaD}
\end{equation}
where
\begin{equation}
    \zeta_{\theta}(s,D)=
{\rm Tr}\left[\left({P_{\theta}(D)\over \mu^2} \right)^{-s}\right]
  = {2\mu^{2s}\over \Gamma(s)}\int_0^{\infty}
      {d\xi\over \xi} \xi^{2s}\
         {\rm Tr}\left[e^{-\xi^2 P_{\theta}(D)}\right]
.
    \label{zepsilon}
\end{equation}
It should be noted that the operator $P_\theta$ is positive and therefore
the integral is well behaved at large $\xi$.


As is well known, the regularized potential $V^D_\theta$ contains a pole
divergence in the limit $D\to 5$. To see that this is the case,
one introduces the asymptotic expansion of the trace for small
$\xi$ \cite{dw,branson,singer},
\begin{equation}
{\rm Tr}\left[f\ e^{-\xi^2 P_{\theta}(D)}\right]
\sim \sum_{n=0}^{\infty} \xi^{n-D}
a^D_{n/2}(f,P_\theta) ,
\label{sdw}
\end{equation}
where $a^D_{n/2}$ are the so-called generalized Seeley-De\,Witt coefficients.
In (\ref{sdw}) we
have introduced the arbitrary smearing function $f(x)$. This is unnecessary for
the present discussion, but it will
be useful later on.
For $n\leq 5$ their explicit form is known for a wide
class of covariant operators, which includes our $P_\theta$.
They are finite and can be
constructed from local invariants (terms constructed from the
metric, the mass term $E_\theta$ and the smearing function $f$), integrated over
spacetime. For even $n$, they receive contributions from
the bulk and from the branes, whereas for odd $n$ they are made
out of invariants on the boundary branes only.

For definiteness, let us focus on the
simplest case of a Dirichlet scalar field, satisfying
\footnote{In \S\ref{sec:confc} we considered an
even parity scalar field, which obeys Neumann  boundary conditions.
In fact, in the conformally trivial case we would get the same result
for an odd
parity field since the spectrum in the flat space
problem is the same.
The only difference is the presence of a zero mode in the even parity case,
which does not contribute to the result (\ref{hattie}).}
\begin{equation}
    \label{dirichlet}
    \chi(z_\pm)=0.
\label{chiboundary}
\end{equation}
We can use the result found in
\cite{klaus,md,vass} to compute the Seeley-De\,Witt coefficients for
a Dirichlet field
with a bulk operator $P=-(\Box+E)$. The lowest order ones for odd $n$
are given by:
\begin{equation}
    \label{ad12}
    a^D_{1/2}(f,P)={-(4\pi)^{(1-D)\over 2}\over4}
    \sum_{i=\pm}\int_{y_i} \sqrt{g_i} f(x) \,d^{D-1}x,
\end{equation}
\begin{eqnarray}
    \label{ad32}
    a^D_{3/2}(f,P)={-(4\pi)^{(1-D)\over 2}\over384}
    \sum_{i=\pm}\int_{y_i} \sqrt{g_i} \,d^{D-1}x
    \Big\{ f
     &\Big(& 96 E  - 16 {\cal R} + 8 {\cal R}_{yy}
    + 7{\cal K}^2 - 10 {\cal K}_{\mu\nu}{\cal K}^{\mu\nu}  \big) + O(f_{;y}, f_{;yy})\Big\}.
\end{eqnarray}
The most relevant for our purposes will be $a^D_{5/2}$:
\begin{eqnarray}
    \label{betaf}
 a^D_{5/2}(f,P)={-(4\pi)^{(1-D)\over 2} \over 5760}
    \sum_{i=\pm}\int_{y_i} \sqrt{g_i} \,d^{D-1}x \Big\{ f
    &\Big(& 720 E^2 - 450 {\cal K}\,E_{;y} + 360 E_{;yy} \nonumber\\
    &+&15 \big( 7{\cal K}^2 - 10 {\cal K}_{\mu\nu}{\cal K}^{\mu\nu}
    + 8 {\cal R}_{yy} - 16 {\cal R}\big) \, E\nonumber\\
    &+& 20 {\cal R}^2 - 48 \,\Box {\cal R} - 17 {\cal R}_{yy}^2
    - 8 {\cal R}_{ab}{\cal R}^{ab}
    + 8 {\cal R}_{abcd}{\cal R}^{abcd} \nonumber\\
    &-& 20 {\cal R}_{yy}\,{\cal R} + 16 {\cal R}_{yy}{\cal R}
    - 10 {\cal R}_{yy}{\cal R}_{yy}
    - 12 {\cal R}_{;yy}     - 15 {\cal  R}_{yy;yy}\nonumber\\
    &-&16 {\cal K}_{\mu\nu}{\cal K}^{\nu\rho} {\cal R}^\mu_\rho
    - 32 {\cal K}^{\mu\nu}{\cal K}^{\rho\sigma}{\cal R}_{\mu\rho\nu\sigma}
    + {215\over8} {\cal K}_{\mu\nu}{\cal K}^{\mu\nu}\,{\cal R}_{yy} \nonumber\\
    &+& 25 {\cal K}_{\mu\nu}{\cal K}^{\mu\nu}\,{\cal R}
    + {47\over2} {\cal K}_{\mu\nu}{\cal K}^\nu_\rho {\cal R}^{\rho}_{~ y \mu y}
    + {215\over16} {\cal R}_{yy}\,{\cal K}^2 \nonumber\\
    &-&{35\over2} {\cal R}\,{\cal K}^2
    -14 {\cal K}_{\mu\nu}{\cal R}^{\mu\nu}\,{\cal K}
    - {49\over4} {\cal K}^{\mu\nu}{\cal R}_{\mu y\nu y}\,{\cal K}
    +42 {\cal R}_{;y}\,{\cal K} \nonumber\\
    &-&{65\over128} {\cal K}^4
    - {141\over32} {\cal K}_{\mu\nu}{\cal K}^{\mu\nu}\,{\cal K}^2
    + {17\over2} {\cal K}_{\mu\nu}{\cal K}^{\nu\rho} {\cal K}^\mu_\rho\,{\cal K}\nonumber\\
    &+& {777\over32} ({\cal K}^{\mu\nu}{\cal K}_{\mu\nu})^2
    - {327\over8} {\cal K}_{\mu\nu}{\cal K}^{\nu\rho} {\cal K}_{\rho\sigma}
    {\cal K}^{\mu\sigma}
    \Big) +O(f_{;y},...,f_{;yyyy}) \Big\}.
\end{eqnarray}
Our notation is as follows.  $E$ is a general scalar function, ${\cal
  R}^{a}_{~bcd}=+\Gamma^{a}_{~bc,d}-\cdots $ is the Riemann tensor, ${\cal
  R}_{bc}={\cal R}^{a}_{~bac}$ is the Ricci tensor and ${\cal R}={\cal
  R}_{ab}g^{ab}$ is the curvature scalar.
The extrinsic curvature is given by
${\cal K}_{\mu\nu}\equiv (1/2)\partial_y g_{\mu\nu}$,
where
$g_{\mu\nu}(y)$ is the induced metric on
$y$-constant hypersurfaces,
and ${\cal K}={\cal K}_{\mu\nu}\,g^{\mu\nu}$.
The vector normal to the boundary is $\partial_y$ so the normal components are
simply the $y$ components.
The $a$, $b, \cdots$ indices run over the extra
coordinate, and over the directions tangential to the branes,
$\mu$, $\nu,\,\cdots$. The omitted terms, represented by
  $O(f_{;y},...)$, are linear combinations
of the derivatives of $f$ with coefficients which depend on ${\cal
  K}_{\mu\nu}$, $E$ and its
derivatives.

As mentioned above, the integral (\ref{zepsilon}) is well behaved for
large $\xi$. For small $\xi$, the integral is convergent for
$2s>D$, as can be seen from the asymptotic expansion (\ref{sdw}).
In the end, we have to consider the limit $s\to 0$, and so we must
keep track of divergences which may arise in this limit. For this
purpose, it is convenient to separate the integral into a small $\xi$
region, with $\xi<\Lambda$, and a large $\xi$ region with
$\xi>\Lambda$, where $\Lambda$ is some arbitrary cut-off.
Substituting (\ref{sdw}) into (\ref{zepsilon}), we can explicitly
perform the integration in the small $\xi$ region for $2s>D$. This gives
\begin{equation}
\zeta(s,D)\sim 2 {\mu^{2s}\over \Gamma(s)}
\left\{\sum_{n=0}^{\infty}
{\Lambda^{n-D+2s}\over n-D+2s} a^D_{n/2}(P_\theta) +
\int_\Lambda^{\infty}
      {d\xi\over \xi} \xi^{2s}
         {\rm Tr}\left[e^{-\xi^2 P_{\theta}(D)}\right]\right\},
\label{wh}
\end{equation}
where we have used the standard notation
$$
a^D_{n/2}(P_{\theta})=a^D_{n/2}(f=1,P_\theta).
$$

The second term in curly brackets is perfectly finite for all values
of $s$. Analytically continuing and taking the derivative
with respect to $s$ at $s=0$, we have
\begin{equation}
\zeta'(0,D)\sim \sum_{n=0}^{\infty}
{2 \Lambda^{n-D} \over n-D} a^D_{n/2}(P_\theta)+ {\rm finite},
\label{dwh}
\end{equation}
where the last term is just twice the integral in (\ref{wh}) evaluated
at $s=0$.
Introducing the regulator $\epsilon = 5-D$,
the ultraviolet divergent part of $V^D_\theta$ is thus given by
\begin{equation}
V_\theta^{div} = -{1\over \epsilon {\cal A}} a^D_{5/2}(P_\theta).
\label{vdiv}
\end{equation}
The divergence is removed by renormalizing the couplings in
front of the invariants which make up the coefficient $a^D_{5/2}$, and
so this infinite term can be dropped. The renormalized
effective potential of
our interest is therefore given by
\begin{equation}
V_\theta=\lim_{D \to 5}\left[V^D_\theta - V^{div}_\theta\right].
\label{recipe}
\end{equation}
To proceed, we need to calculate $V^D_\theta$, which in principle requires
calculating a trace which involves the eigenvalues of $P_\theta$, and
as mentioned above, these are not related in any simple way to the
KK masses.

However, it turns out that the dimensionally regularized
$V^D_\theta$ is independent of $\theta$ when
$D$ is not an integer.
The dependence of $V_\theta^D$ on $\theta$
can be found in the following way.
First we note that
\begin{equation}
\partial_\theta{\rm Tr}\left[e^{-\xi^2 P_\theta}\right] =
{\rm Tr}\left[2\xi^2 f_\theta(x) \Omega_\theta^{-2}
P e^{-\xi^2 P_\theta}\right]=-\xi\partial_\xi
{\rm Tr}\left[f_\theta(x) e^{-\xi^2 P_\theta}\right],
\label{rel2}
\end{equation}
where we have introduced
$$
f_\theta\equiv \partial_\theta\ln\Omega_\theta,
$$
and the cyclic property of the trace was used.
The above relation enables us to find the dependence of
$V^D_\theta$ on the conformal factor:
\begin{equation}
\partial_\theta \lim_{s\to 0}\partial_s \zeta_\theta(s,D) =
\lim_{s\to 0} \partial_s {2\mu^{2s} \over \Gamma(s)}
\int_0^{\infty} d\xi \xi^{2s} \partial_\xi {\rm Tr}\left[
-f_\theta e^{-\xi^2 P_\theta}\right].
\end{equation}
As with the expansion (\ref{wh}) we may again
introduce the regulator $\Lambda$ and separate the
integral into a large $\xi$ part with $\xi>\Lambda$,
which is finite and a small $\xi$ part with $\xi<\Lambda$
which contains the divergent ultraviolet behaviour.
Assuming that $2s>D$ and integrating by parts, the resulting integrals in the small $\xi$ region can
be performed explicitly and we have
\begin{equation}
\partial_\theta \lim_{s\to 0}\partial_s \zeta_\theta(s,D) \sim
\lim_{s\to 0} \partial_s
{4 s \mu^{2s} \over \Gamma(s)} \left[
\sum_{n=0}^{\infty}
{\Lambda^{n-D+2s}\over n-D+2s} a^D_{n/2}(f_\theta,P_\theta)
+{\rm finite}\right].
\label{derivative}
\end{equation}
As before, the last term just indicates the integral in the large
$\xi$ region.
Provided that $D$ is not an integer, all terms in square brackets
remain finite at small $s$, and so the right hand side of
(\ref{derivative})
vanishes.
Hence, we find that
\begin{equation}
\partial_\theta V^D_\theta = 0, \quad (D\neq {\rm integer}).
\label{invth}
\end{equation}
In other words,
the dimensionally regularized
determinant of $P_\theta$ coincides with the dimensionally regularized
determinant of $P_0$, and we have
\begin{equation}
V^D_\theta = V_0^D \equiv V^D \equiv
\sum_n \mu^{\epsilon}{1\over 2}\int {d^{D-1}
      k\over
      (2\pi)^{D-1}}\log\left({k^2 + m_n^2(\varphi_i,D)
      \over\mu^2}\right),\quad \quad (D\neq {\rm integer}).
\label{nafta}
\end{equation}
As was anticipated, we find
that $\ln J_{\theta}$ vanishes
in the dimensional regularization presented here in the
sense given in (\ref{TrlnOmega}).

Finally, from (\ref{recipe}) and (\ref{nafta}), the renormalized
effective potential is given by
\begin{equation}
V_\theta(\varphi)= \lim_{D \to 5} \left[ V^D - {1\over (D-5)} {1\over\
{\cal A}}\
a^{D}_{5/2}(P_\theta) \right],
\label{true}
\end{equation}
where the Seeley-De\,Witt coefficient $a^{D}_{5/2}$ is given in
(\ref{betaf}) with $f=1$.
The above equation bears the ambiguity in the choice of integration
measure in the second term in square brackets. Different values of
$\theta$ give different results. If we take $g_{ab}$ as the preferred
metric, then we should use
$\theta=1$, whereas if we take $g^{(s)}_{ab}$ as the preferred metric,
we should use $\theta=-1/\beta$. As we shall see in the next
subsection, when we set $D=5$ the coefficient $a_{5/2}(P_\theta)$ is also
independent of $\theta$. Hence, the pole term in the second term in
(\ref{true}) is independent of $\theta$, as it should, in order to cancel
the pole in $V^D$. However, the finite part does depend on the choice
of $\theta$.

The right hand side of (\ref{true}) is ready for explicit evaluation,
which is deferred
to the next section. We shall now turn our attention to the equivalent method
of zeta function regularization.


\subsection{Zeta function regularization}

The method of zeta function regularization exploits the fact that
the formal expression for the effective potential (\ref{vthetaD}) is
finite if the limit $D\to
5$ is taken before the limit $s\to 0$. This can be seen from Eq.
(\ref{wh}), where the term with $n=5$ is finite
if we set $D=5$ before taking the derivative with respect to $s$
and setting $s\to 0$. Clearly, the change in the order of the limits
simply removes the divergent term $V^{div}$ given in (\ref{vdiv}) and it
reproduces the results obtained by the method of dimensional
regularization (up to finite renormalization terms which are
proportional to the geometric invariant $a_{5/2}^{D=5}(P_\theta)$).

In zeta function regularization we define
\begin{equation}
    V_{\theta}\equiv -{1\over 2{\cal A}}
    \ln(\det P_\theta) \equiv -{1\over 2 {\cal A}}\lim_{s\to 0}\partial_{s}
    \zeta_{\theta}(s),
    \label{vtheta}
\end{equation}
where $\zeta_{\theta}(s)\equiv \zeta_{\theta}(s,D=5)$
[see Eq. (\ref{zepsilon})].
As in the case of dimensional regularization,
it is more convenient to calculate $V_0$ than $V_\theta$
since the eigenvalues of $P_0$ are related to the spectrum of KK masses.
An important
difference with dimensional regularization is that
$$
-2 {\cal A}\partial_\theta V_\theta = \partial_\theta\zeta'_\theta(0)=2a_{5/2}(f_\theta,P_\theta)\neq 0,
$$
a result which we already encountered in Ref. \cite{gpt} (see also \cite{klaus}).
This can be seen from (\ref{derivative}). If we set $D=5$ from the
very beginning, the term with $n=5$ in Eq. (\ref{derivative})
is linear in $s$, and its derivative with respect to $s$
does not vanish in the limit $s\to 0$.
Here, and in what follows, we use the notation
$$
a_{n/2}\equiv \lim_{D\to 5} a_{n/2}^D.
$$
Integrating along the conformal path parameterized by $\theta$, we
can relate the effective potential per unit comoving volume $V_\theta$,
with the "flat space" effective potential $V_0$ as
\begin{equation}
    V_\theta=V_{0}-{1\over {\cal A}}
    \int_0^\theta d\theta'\, a_{5/2}(f_{\theta'},P_{\theta'}).
    \label{conformal}
\end{equation}
The general expression for $a_{5/2}(f_\theta,P_\theta)$ which
applies to our case has been derived by Kirsten\cite{klaus}.
In Ref. \cite{gpt} we evaluated the integral in (\ref{conformal})
for the Randall-Sundrum case, in order to obtain $V_1$ from $V_0$.
Here we shall present an alternative expression for this integral
which does not require the knowledge of
$a_{5/2}(f_\theta,P_\theta)$,
but only the knowledge of $a_{5/2}^D(P_\theta)$ for dimension $D=5-\epsilon$.
This will also illustrate the
relation between the method of zeta function regularization
and the method of dimensional regularization.

From the asymptotic expansion of the first and the last terms in
Eq. (\ref{rel2}), we have\cite{klr}
\begin{equation}
\partial_\theta a^D_{n/2}(P_\theta) = (D-n) a_{n/2}^D(f_\theta,
P_\theta).
\label{oriol}
\end{equation}
Integrating over $\theta$, we get
\begin{equation}
    (D-5) \int_0^\theta  a^{D}_{5/2}(f_{\theta'},P_{\theta'}) d\theta'
    =a^{D}_{5/2}(P_\theta)-a^{D}_{5/2}(P_0) .
\end{equation}
Writing $D=5-\epsilon$, we have
\begin{equation}
 V_\theta-V_0 =  - {1\over {\cal A}}
 \int_0^\theta    a_{5/2}(f_{\theta'},P_{\theta'}) d\theta'
    ={1\over \epsilon{\cal A}}\left[
    a^{D}_{5/2}(P_\theta)-a^{D}_{5/2}(P_0)\right].
\label{integ}
\end{equation}
Note that,
from (\ref{vdiv}) and (\ref{true}), the previous equation
can also be written as
\begin{equation}
V^D \equiv V_\theta + V^{div}_\theta = V_0 + V^{div}_0.
\end{equation}
This equation simply expresses the fact that
the dimensionally regularized $V^D$ is independent of the conformal
parameter $\theta$, as we had shown
in the previous Subsection [see e.g. Eq. (\ref{invth})].

From (\ref{oriol}), with $D=n=5$, one finds that the coefficient
$a_{5/2}(P_\theta)$ is conformally invariant \cite{klr}, and therefore
\begin{equation}
a_{5/2}(P_\theta)=a_{5/2}(P_0).
\label{sempervivens}
\end{equation}
Substituting this into
(\ref{integ}), we obtain
\begin{equation}
    \int_0^\theta    a_{5/2}(f_{\theta'},P_{\theta'}) d\theta'
    ={d\over d  D} a^{D}_{5/2}(P_\theta)\Big|_{D=5}-
       {d\over d  D} a^{D}_{5/2}(P_0)\Big|_{D=5}.
\end{equation}
Thus, the integral in (\ref{conformal}) can
be evaluated in two different ways. One is
by using the explicit expression of $a_{5/2}(f,P)$
given by Kirsten\cite{klaus}. The other is by taking the derivative of the
coefficients $a^D_{5/2}(P_\theta)$, given in (\ref{betaf}) with $f=1$,
with respect to the dimension. [Note that the
terms which are linear in derivatives of $f$, which we have just
indicated symbolically in (\ref{betaf}), disappear when $f$ is a constant].

%
%
%

The quantity discussed above is
nothing but the $\theta$ dependence induced by the choice of
integration measure. In the sense of (\ref{TrlnOmega}), we have
\begin{equation}
  {\ln} J_\theta
      = {1\over 3c {\cal A}}
    \int^1_\theta d\theta'\, a_{5/2}(
        \phi,P_{\theta'}),
\label{onu}
\end{equation}
where we have used $f_\theta = \partial_\theta \ln\Omega_\theta
=(1-\theta)\phi/3c$.
Clearly,
the effect of this factor is just adding local terms
expressed solely in terms of $\phi$ and the metric
to the classical action. The dependence of these terms
is different from the change which results
from a rescaling of the renormalization parameter $\mu$. This corresponds to
a shift in the coefficient of local terms proportional to $a_{5/2}(P)$.

\section{Explicit evaluation}

For simplicity we shall restrict attention to
the case of massless fields with arbitrary coupling to the curvature:
$$
E=-\xi {\cal R}_g,
$$
and with Dirichlet boundary conditions. Here we shall use
the method of dimensional regularization. Zeta function regularization
is discussed in the Appendix.

The eigenmodes in
(\ref{massess}) are given by
$$
\hat\chi_n\equiv
 z^{1/2}\big( A_1 J_\nu (m_n z) +A_2 Y_\nu (m_n z)\big).
$$
The index of the Bessel functions is given by
\begin{equation}
    \label{index}
    \nu(D)={1\over2}\sqrt{1-4 (D-1)\beta[(D-2)\beta-2](\xi-\xi_c(D))},
\end{equation}
where
$$
\xi_c(D)={1\over 4}{D-2\over D-1},
$$
is the conformal coupling in dimension $D$.
Imposing the  boundary conditions (\ref{dirichlet}) on both branes,
we obtain the equation that defines implicitly the discrete spectrum
of $m_n$,
\begin{equation}
    F(\tilde m_n)=J_{\nu}(\tilde m_n \eta)Y_{\nu}(\tilde m_n)
    -Y_{\nu}(\tilde m_n \eta)J_{\nu}(\tilde m_n)=0,
    \label{mspectrum}
\end{equation}
where we have defined
\begin{equation}
\tilde m_n= m_n z_-, \quad \quad \eta= {z_+\over z_-}.
\label{tildem}
\end{equation}
The zeros of $F$ are all real if $\nu$ is real and $\eta$ is positive.  Since
$\eta$ is positive, the reality condition of $\nu$ guarantees that all the KK
masses are real. This provides a constraint for the possible values
of
$\xi$ depending on ${\kk}$,
\begin{eqnarray}
    \label{allowedxi}
    &&\displaystyle \xi\geq{-(1-4{\kk})^2\over 16{\kk}(2-5{\kk})},  \qquad
    {\kk}\leq2/5, \cr
    &&\displaystyle\xi\leq{-(1-4{\kk})^2\over 16{\kk}(2-5{\kk})},
    \qquad{\kk}\geq2/5,
\end{eqnarray}
where we have used $D=5$.
Note that the values of $\xi$
comprised between the minimal and the conformal coupling are allowed for any
value of ${\kk}$.

\begin{figure}[htbp]
    \begin{center}
        \epsfysize=7 cm \epsfbox{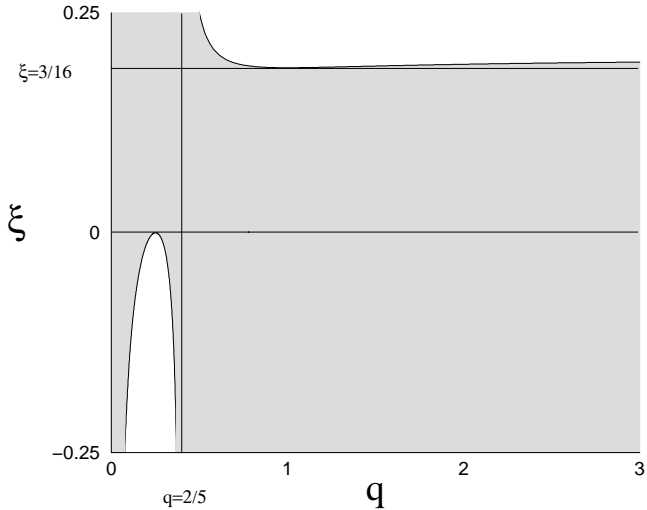}
               \label{fig:xikappa}
               \caption{\footnotesize The shaded region corresponds to the
                 allowed values in the ($\kk$ --$\xi$) plane for a
Dirichlet massless scalar
                 field according to Eq. (\ref{allowedxi}). Note that the range
                 of $\xi$ comprised between the minimal and the conformal
                 coupling is allowed for any value of $\kk$}
    \end{center}
\end{figure}

In section V we concluded that the renormalized expression for
the effective potential is
\begin{equation}
V_\theta(\varphi)= \lim_{D \to 5} \left[ V^D - {1\over (D-5)} {1\over\
{\cal A}}\
a^{D}_{5/2}(P_\theta)  \right].
\end{equation}
Consider first  $V^D$, given in Eq. (\ref{4b}).
Performing the momentum integrations, we obtain
\begin{equation}
    \label{unrenpot}
    V^D=
    - {1\over 2(4 \pi)^2}(4\pi\mu^2)^{\epsilon/2}
    {1\over z_-^{4-\epsilon}}
   \Gamma(-2+\epsilon/2) \tilde\zeta(\epsilon-4),
\end{equation}
where we have defined \cite{gpt}
\begin{equation}
   \tilde\zeta(s)= \sum_n \tilde m_n^{-s}=
   {s\over 2\pi i}\int_{\cal C} t^{-1-s}\ln F(t)\ dt.
\label{zh}
\end{equation}
This regularized expression
for the effective potential is finite when the real part of $\epsilon$ is
sufficiently large.
In the last equation we have used that $F(t)$ has only simple
zeros which are along the real axis. The closed contour of integration
${\cal C}$ runs along the imaginary axis, from $t=+i\infty$ to $t=-i\infty$,
skipping the origin through an infinitesimal path which crosses the
positive real axis, and the contour is closed at infinity also
through positive real infinity.

Now the problem reduces to the computation of $\tilde \zeta$,
which can be done in the same way as in the case discussed in \cite{gpt}.
Skipping the detailed derivation, we
simply give the final result:
\begin{equation}
    \label{tildezeta}
    \tilde\zeta(-4+\epsilon)
    =-2\beta_4(1+\eta^{-4+\epsilon})-2\epsilon
        (\eta\, \tau)^{-2} \Big({\cal I}_K  +{\cal I}_I \tau^4  +
            \tau^4 {\cal V}(\tau)
        \Big)
    +O(\epsilon^2),
\label{skippy}
\end{equation}
where
\begin{equation}
\beta_4={1\over128}(13-56\nu^2+16\nu^4).
\label{beta4}
\end{equation}
Here we have introduced
$$
\tau\equiv {z_<\over z_>} = \left\{
    \begin{minipage}[c]{5cm}
    $1/\eta,      \quad \text{for} \quad {\kk}<1$,\newline
    $~\eta,     \quad \text{for} \quad {\kk}>1$,
    \end{minipage}
\right.
$$
to express the result for both $q>1$ and $q<1$ cases simultaneously,
where $z_>$ and
$z_<$ are the largest and the smallest of $z_+$ and $z_-$, respectively.
Note that $\tau<1$.

\begin{figure}[htbp]
    \begin{center}
        \epsfxsize=14 cm \epsfbox{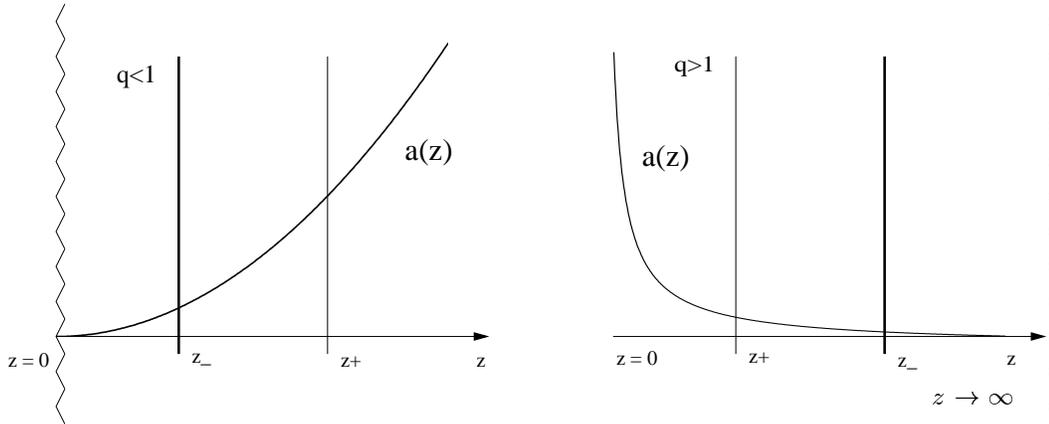}
        \label{fig:confcoord}
        \caption{\footnotesize For ${\kk}<1$, in conformal coordinate, the
          singularity sits at $z=0$, and the coordinate of the negative tension
          brane is smaller than the coordinate for the positive tension one,
          $\eta=\ z_+/z_-\ >1$.  For ${\kk}>1$, the singularity is placed at $z\to\infty$
          and $z_- > z_+$, so $\eta<1$.}
    \end{center}
\end{figure}

The constant coefficients ${\cal I}_K(\nu)$, ${\cal I}_I(\nu)$ are
calculable in principle, although their precise value is perhaps
not very interesting since as we shall see these coefficients can be
reabsorbed by finite renormalization. Their definition is given in
the Appendix. Finally, ${\cal V}(\tau)$ is
defined by
\begin{equation}
{\cal V}(\tau)=\int_0^{\infty} d\rho\,\rho^{3}
         \ln\left(1-{I_{\nu}(\tau\rho)\over K_{\nu}(\tau\rho)}
                 {K_{\nu}(\rho)\over I_{\nu}(\rho)}\right).
\label{calvo}
\end{equation}

This provides the following expression for $V^D$
\begin{eqnarray}
    V^D&=&{1\over (4\pi)^2}
    \Biggr[
        \left\{\Big({1\over\epsilon}+{3\over4}-{\gamma\over2}
            +{1\over2} \ln(4\pi \mu^2 z_0^2) \Big)
            \beta_4
            +\beta_4'\right\}
        \left({1\over z_-^4}+{1\over z_+^4}\right) \\
        &+&\beta_4
        \left({1 \over z_-^4}\ln \Big({z_-\over z_0}\Big)
            +{1 \over  z_+^4} \ln \Big({z_+\over z_0}\Big)\right)
        +{{\cal I}_K \over z_<^4}+ {{\cal I}_I\over z_>^4} +
        {{\cal V}(\tau)\over z_>^4}\Biggr] +{\cal O}(\epsilon),
    \label{flatstresultDR}
\end{eqnarray}
where we have introduced the conventions
$
\beta_4\equiv\beta_4(D=5)
$ and
$
\beta_4'\equiv {d\beta_4(D)/dD}|_{D=5},
$
since $\beta_4$ depends on the dimension through (\ref{index}).

The next step is to subtract the divergent contribution.
The Seeley-De\,Witt coefficient $a_{5/2}^D(P_\theta)$ can be
computed from (\ref{betaf}) for a generic dimension
$D$, so we can expand the second term in the r.h.s of Eq. (\ref{true})
as
\begin{equation}
    {1\over (D-5)} {1\over {\cal A}} a^{D}_{5/2}(P_\theta) =
{1\over (4\pi)^2}\left[
        {1\over \epsilon}\beta_4
        \left({1\over z_-^4}+{1\over z_+^4}\right)
        -\beta\theta \;\beta_4
        \left({1 \over z_-^4}\ln \Big({z_-\over z_0}\Big)
            +{1 \over  z_+^4} \ln \Big({z_+\over z_0}\Big)\right)
        -\delta(\theta)\left({1\over z_-^4}+{1\over z_+^4}\right)
    \right],
    \label{divpartDR}
\end{equation}
where $\delta(\theta)$ is a constant whose precise value will be unimportant.
\footnote{For $\theta=1$, we find
$\delta=(-\beta/3072)\{432 - 1936\beta - 900\beta^2 + 2257\beta^3
- 1152(2
+ 7(-1 + \beta)\beta(1 + 3\beta))\xi + 3072\beta(-2 + 3\beta)(-2 +
7\beta)\xi^2\}$.}

The divergent parts of the two terms in Eq.
(\ref{true}) cancel, and the finite result for the effective potential
per unit comoving volume is given by
\footnote{We can check
  that this coincides with Eq.  (\ref{hattie}) in the conformal case,
  $\nu=1/2$ both for the Dirichlet or Neumann boundary conditions.  The
  trace anomaly of conformally coupled fields in this conformally flat
  background is zero for this topology and for any kind of boundary conditions,
  and so  $\beta_4=0$.  Moreover, if $\nu=1/2$ then $K_{1/2}=\sqrt{\pi/2
    \rho}\,e^{-\rho}$ and $I_{1/2}=\sqrt{2/ \pi \rho}\,\sinh\rho$. This means
  that ${\cal I}_K$, ${\cal I}_I$ and ${\cal V}(\eta)$ can be computed
  analytically, easily recovering the result Eq.~(\ref{hattie}).
}
\begin{equation}
    \label{veffnonmin}
    V_\theta={1\over (4\pi)^2}
    \left[{{\cal I}_ K \over z_<^4}+ {{\cal I}_I\over z_>^4} +
    {{\cal V}(\tau)\over z_>^4}\right]
    + {\beta_4 (\beta\theta +1) \over (4\pi)^{2}}
    \left[{1\over z_+^4} \ln\left({z_+\over z_0}\right) +{1\over z_-^4 }
    \ln\left({z_-\over z_0}\right)\right]
\end{equation}
$$
    + {\beta_4 \over (4\pi)^{2}}
    \left[{1\over z_+^4}+ {1\over z_-^4 }\right]\ln(\mu z_0).
$$
Here, we have elliminated some terms through redefinition of $\mu$.

As in the case of conformal fields, we must also allow for finite
renormalization of the couplings in front of the invariants
which make up the coefficient $a_{5/2}$. These have the general
dependence of the form $a_{\pm}^4 {\cal K}_\pm^4 \sim z_\pm^{-4}$.
With these additions, we finally obtain:
\begin{equation}
    \label{veffnonmin2}
    V_\theta(z_+,z_-)=
     {\beta_4 (\beta\theta+1) \over (4\pi)^{2}}
     \left[{\ln\left(\mu_1{z_+}\right)\over z_+^4} +
    {\ln\left(\mu_2{z_-}\right)\over z_-^4}\right]
    + {1\over (4\pi)^2}
\int_0^{\infty} dx\,x^3
         \ln\left[1-{I_{\nu}(x z_<)\over I_{\nu}(x z_>)}
                 {K_{\nu}(x z_>)\over K_{\nu}(x z_<)}\right]
    ,
\end{equation}
where $\mu_1$ and $\mu_2$ are renormalization constants, and
$\beta$ and $\beta_4$ are given in equations (\ref{defz0}) and
(\ref{beta4}). This is the main result of this Section.
In the limit of small separation between the branes, $(1-\tau)\ll 1$,
the integral ${\cal V}$ behaves like $(1-\tau)^{-4}$ (see Appendix C), and the
logarithmic terms can be neglected. In this limit, the potential
behaves like the one for the conformally coupled case, given in
(\ref{ve1}):
\begin{equation}
    V_\theta(z_+,z_-) \sim - {A \over |z_+ - z_-|^4}.
\end{equation}
For $\tau\ll 1$, when the branes are well separated,
the integral ${\cal V}$
behaves like $\tau^{2\nu}$ and becomes negligible in the
limit of small $\tau$ (except in the special case when $\nu$ is
very close to 0). In this case, we have
\begin{equation}
V_\theta(z_+,z_-) \sim
{\beta_4 (\beta\theta+1) \over (4\pi)^{2}}
     \left[{\ln\left(\mu_1{z_+}\right)\over z_+^4} +
    {\ln\left(\mu_2{z_-}\right)\over z_-^4}\right] +
O\left[\left({z_< \over z_>}\right)^{2\nu}\right].
\label{logs}
\end{equation}
Due to the presence of the
logarithmic terms, it is in principle possible to adjust the
parameters $\mu_1$, $\mu_2$ so that there
are convenient extrema for the moduli $z_+$ and $z_-$.
We shall comment on this point in the concluding section.

\section{An Example}
\label{sec:hw}

Here we shall consider an example of a contribution to the moduli effective
potential in the model of Lukas et al.  \cite{ovrut,hw1,hw2},
which may be relevant for the Ekpyrotic universe \cite{ekpy}.
As mentioned above, this corresponds to the case $q=1/6$. In principle,
the fields in the action for this model do not have the canonical form, since
in addition to the coupling to the metric they have unusual couplings
to the dilaton $\phi$. Nevertheless, they can be
studied along the lines of the previous Sections. For instance,
the Heterotic M-theory model of \cite{ovrut,hw1,hw2}
contains a scalar $\Psi(x^\alpha)$ in the universal hypermultiplet
whose vev is zero in the background solution, and whose action is
given by
\begin{equation}
   S^{(\Psi)} = -\int \sqrt{g} \, d^5 x
        \;\frac{1}{2} e^{-\phi} (\partial \Psi)^{2} .
\label{xiaction}
\end{equation}
This contains a kinetic term only, but it has a nonminimal form.
However, changing to a new variable,
$$
\varphi=e^{-\phi/2} \Psi,
$$
we rewrite the action (\ref{xiaction}) as
\begin{eqnarray}
   S^{(\Psi)} &=& -\int \sqrt{- g} \, d^5 x \nonumber
   \;\frac{1}{2}\Big(\partial\varphi +{1\over2}\varphi\,\partial\phi\Big)^2\\
   &=& -\int \sqrt{- g} \, d^5 x
   \;\frac{1}{2} \Big( \partial\varphi^2-      \nonumber
   {1\over2}\,\big(\Box \phi -{1\over2}\,(\partial\phi)^2\big)\,\varphi^2 \Big)
   +\text{boundary terms}\\
   &=& -\int a^4 \, d^4x \,dy
   \;\frac{1}{2} \varphi \, \Big( -\Box -
   {5\over12}\,{1\over y^2} \Big) \,\varphi,
\end{eqnarray}
where the indices are contracted with the five dimensional
$g_{ab}$ metric, and we have
integrated by parts in the first equality. For the Dirichlet case,
the boundary terms 'generated' are not relevant since still the field
$\varphi$ vanishes there.  The last equation shows that
the potential term present in
terms of $\varphi$ mimics a nonminimal coupling to the
curvature $-\xi{\cal R}\,\varphi^2$.
Since the Ricci scalar for this background
is
$$
{\cal R}={7\over9}\,{1\over y^2},
$$
we conclude that the equivalent effective nonminimal coupling of $\varphi$ is
$\xi=15/28$. Note that this point lies in the allowed region of
values in the ($\kk$ --$\xi$) plane defined by Eq. (\ref{allowedxi}), which
corresponds to an index of the Bessel functions $\nu=4/5$.

Hence, the contribution to the moduli effective potential induced by the
field $\Psi$ is given by (\ref{veffnonmin2}) with $\nu=4/5$. In this case
we have $\beta=1/5$ and $\beta_4=-10179/80000$. There are, of course, many
other contributions,
corresponding to all bosonic and fermionic degrees of freedom. It turns out,
however, that all bosonic contributions take a form similar to the one
of the field $\Psi$ \cite{gpt3}. If supersymmetry is unbroken, then these
contributions are cancelled by the contributions from the fermionic
degrees of freedom. But
if the degeneracy between bosons and fermions is broken a la
Scherk-Schwartz, for instance, then we expect that the resulting effective
potential will be qualitatively similar to the one given by
(\ref{veffnonmin2}). A detailed study of this model is left for future
research.

\section{Moduli stabilization}

In the limit of large interbrane separation, the potential
(\ref{veffnonmin2}) assumes a ``Coleman-Weinberg'' form for each one of
the moduli,
\begin{equation}
V(y_+,y_-) \approx
\sum_{i=\pm}
 \ a^4(y_i) \left\{\alpha
 K^4(y_i)\ln\left[{K(y_{i})\over \mu_i}\right] + \delta\sigma_i \right\}.
\label{sumsup}
\end{equation}
Here, we have expressed the potential in terms of the "curvature scale"
$$
K(y) = {q\over y},
$$
so that $K^4(y_i)$ behaves like a generic geometric operator of
dimension $4$ on the brane [such as the fourth power of the
extrinsic curvature, or any of the operators in the integrand of
Eq. (\ref{betaf})]. Working with $K_i$ instead of $z_i$ has the advantage
of relating directly to physical quantities, and hence it is easier to
control whether we are in the range where the effective theory should be
trusted or not. In particular, we should not
allow $K_i$ to  be bigger than the cut-off scale of the theory.
The
constant $\alpha$ in (\ref{sumsup}) is given by
\begin{equation}
\alpha={(1-\theta)q-1 \over (4\pi)^{2}}\
(1-q^{-1})^4\sum\beta^{(\chi)}_4,
\label{alphaeq}
\end{equation}
where we sum over the contributions from all bulk fields $\chi$.
The numerical coefficients
$\beta^{(\chi)}_4$ are given by Eq. (\ref{beta4}).
The value of $\theta$ depends on the
choice of integration measure in the path integral which defines the
effective potential (see Section IV).
If we adopt the measure associated to the Einstein frame metric $g_{ab}$
which enters our original action functional (\ref{general}), then
we should take $\theta=1$. However, this is not the only
possible choice, as we have repeatedly emphasized. The clasical action
has a scaling symmetry which transforms both $g_{ab}$ and the
background scalar field $\phi$. Using a conformal
transformation which involves the scalar field, we may construct a
new metric $g^{(s)}_{ab}$ which does not transform under scaling.
If we adopt the measure which corresponds to this new metric, then we should take
$\theta=1-1/q$. With this particular choice of $\theta$ the coefficient
$\alpha$ vanishes and the logarithmic terms in (\ref{sumsup})
disappear. Nevertheless, it is far from clear that this
is indeed a preferred choice
\footnote{In particular,
as mentioned in Section II, for certain values of the
model parameters our action can be obtained by dimensional reduction
of a 5+n dimensional Einstein-Hilbert action with a cosmological
term. Since in the
higher dimensional theory only gravity is present, there is only one
possible choice for the metric and the issue of the measure does not
arise. Preliminary analysis of the case with n=1 \cite{gpt3} indicates
that the logarithmic terms do indeed arise in the limit when the
radius of the sixth dimension is much smaller than the interbrane
separation. For this case the relevant value of the parameter is $\theta=0$.}
Here we take the attitude that the
parameter $\theta$ is unknown, and that it should be fixed by a more
fundamental theory of which (\ref{general}) is just a low energy limit.

The renormalization constants $\mu_i$ in (\ref{sumsup}) can be estimated by
looking at the "renormalized coefficient" of the geometric terms
of dimension $4$ on the brane
$c_i(K) = \alpha \ln(K/\mu_i)$.
In the absence of fine-tuning, the $c_i(K)$ are expected to be of order one
near the cut-off scale $K \sim M$, where $M^{-3}$ is basically the
five-dimensional Newton's constant. Hence, we expect
\begin{equation}
\mu_i \sim M e^{-c_i/\alpha},
\label{muy}
\end{equation}
where $c_i=c_i(M)\sim 1$. In (\ref{sumsup}), we have also allowed for finite
renormalization of local operators on each one of the branes. These
operators are collectively denoted by
$\delta\sigma_i$. In order to
ensure that the effective potential $V$ does not severely distort
the background solution, this correction to the brane tension must
be much smaller than the effective tension of the brane in the
classical background solution. From the Darmois-Israel matching
conditions, this effective tension is of order $M^3 K_i$.
Hence we require
\begin{equation}
\delta\sigma_i \ll M^3 K_i \ll M^4.
\label{dskm3}
\end{equation}
In Section IV we considered contributions
to the effective potential from massless bulk fields. These
may have an arbitrary coupling to the curvature scalar of
the standard form $\xi {\cal R} \chi^2$, or certain couplings
to the background scalar field, such as the ones occurring in the
Heterotic M-theory
model considered in Section VI. However, if the model contains massive
bulk fields, of mass $m$, then we expect terms proportional to
$m^2 K^2$ in the effective potential. Even without massive
bulk fields, we may expect the presence of lower dimensional
worldsheet operators of the form $M^3 K$, $M^2 K^2$ and $M K^3$,
due to cubic, quadratic and linear divergences in the effective
theory. Hence, we may expect that $\delta\sigma_i$ has an expansion of the form
\begin{equation}
\delta\sigma_i(K_i) \sim \Lambda_i^4 + \gamma_{1i} M^3 K_i +
\gamma_{2i} M^2 K_i^2 + \gamma_{3i} M K_i^3+ \gamma_{4i} K_i^4 + {\cal O} (K_i^5),
\label{expansion}
\end{equation}
where $K_i\ll M$, $\Lambda_i\ll M$ and $\gamma_{1i} \ll 1$ in order to satisfy
(\ref{dskm3}). For completeness, the above expansion includes the term proportional
to $K_i^4$. It should be understood that this term is only present in the
particular case $\theta=1-1/q$ (corresponding to the scale invariant metric),
since for other values of $\theta$ we assume that
it is reabsorbed in a redefinition of $\mu_i$ [see Eq. (\ref{muy})].

The local terms may in principle stabilize the moduli at
convenient locations. Note that this effect is due to the warp factor
and vanishes in flat space (where the coefficients $\beta_4$ vanish).
The effect also vanishes accidentally in the RS case, because the
curvature scale $K(y)$ is constant.
The position of the minima are determined by
$
{\partial_{y_i} V} =0.
$
This leads to the conditions
\begin{equation}
\delta\sigma_i = {\alpha\over q}K_i^4
\left[(1-q)\ln\left({K_i\over\mu_i}\right)+{1\over 4}\right]+
K_i {\delta\sigma_i'\over 4q},
\label{extremum}
\end{equation}
where the prime on $\delta\sigma_i$ indicates derivative with respect
to $K_i$.
Also, we must require that the minima occur at an acceptable value
of the effective cosmological constant. Using the condition
(\ref{extremum}), we can write the value of the potential at the
minimum as

\begin{equation}
V_{min}={K_+^4\over 4q}\sum_{i=\pm} \left({K_i\over K_+}\right)^{4(1-q)}
\left\{4\alpha \ln\left({K_i\over\mu_i}\right)+ \alpha + K_i^{-3}
\delta\sigma'_i\right\}
\lesssim 10^{-122} m_p^4.
\label{ccp}
\end{equation}
The latter condition will require one fine tuning amongst the
parameters in (\ref{expansion}).

An interesting question is whether the effective potential (\ref{sumsup})
can generate a large hierarchy and at the same time give sizable masses
to the moduli. As discussed in Section II.B, the hierarchy is given by
\begin{equation}
h^2 = {m^2 \over m_p^2} \sim {K_+\over M} \left({K_+\over
K_-}\right)^{2q-2},
\label{hierarchy}
\end{equation}
where $m\sim TeV$ is the mass of the particles which live on the negative
tension brane, as perceived by the observers on the positive tension
brane. Consistency
with Newton's law at short distances requires
$K_+ \gtrsim (TeV)^2/m_p \sim (mm)^{-1}$, and consistency of perturbative
analysis requires $K_-\lesssim M$. With these constraints,  the observed
hierarchy $h\sim \exp(-37)$ can only be accomodated for
$q\gtrsim 5/4$. To proceed, we shall distinguish two different
cases.

\subsection{Case $a$:}

This is the generic case, where the coefficients
$\gamma_{1i}$, $\gamma_{2i}$ and $\gamma_{3i}$ in the expansion of
$\delta\sigma_i(K)$ are not too suppressed. In this case,
the logarithmic terms in the effective potential are in fact
subdominant, and the minima of the effective potential are
determined by $4q \delta\sigma_i \approx K_i\delta\sigma'_i$.

The present discussion applies also to the special case where $\theta=1-1/q$
(corresponding to the measure associated with the scale invariant metric
$g^{(s)}_{ab}$), so that no logarithmic terms are present in the effective potential.
Note that terms of the form $\gamma_{4i} K_i^4$ and $\gamma_{1i} M^3 K_i$
in the expansion of $\delta\sigma_i(K)$ [see (\ref{expansion})] cannot be avoided.
The first is necessary in order to renormalize the effective potential, and the
second is already present at the tree level, so it just corresponds to a
shift in the existing parameters in the classical action.

Quite generically, this will lead to stabilization of the moduli near
(or slightly below) the cut-off scale $K_i = \lambda_i M$, with
$\lambda_i \sim 1$.
Hence
we have
$$
h^2\sim \exp[2(q-1) \ln(\lambda_+/\lambda_-)].
$$
Since the logarithm is of order one, an acceptable hierarchy can
be generated provided that
$
q \gtrsim 10.
$
This is "close" to the RS limit
$
q\to \infty.
$
In this case, $m_p \sim M$. On the
positive tension brane the parameter $\Lambda_+$ has to be fine
tuned so that the effective cosmological constant is 122 orders of
magnitude smaller than the Planck scale.
A straightforward calculation shows that the physical mass eigenvalues
for the moduli $\varphi_{\pm}$ in the present case are given by
$$
m_+^2 \sim q^{-2}{m_p^{-2}K_+^4}\lesssim m_p^2\,,\quad \quad
m_-^2 \sim q^{-1} h^2 {m_p^{-2}K_-^4}\lesssim h^2 m_p^2.
$$
Thus, the lightest radion has a mass comparable to the $TeV$
scale.

\subsection{Case b:}

This corresponds to the case where almost all of the
operators in (\ref{expansion}) are either extremely suppressed or
completely absent, due perhaps to some symmetry. In particular,
we shall concentrate on the possibility that
$$
\delta\sigma_i = \gamma_{1i} M^3 K_i,
$$
since an operator of this form is already present in the
classical action (\ref{general}), and it is the only one in the
expansion (\ref{expansion}) which is allowed by the scaling symmetry.
In this case, and assuming for simplicity that the
negative tension brane is near the cut-off scale $K_- \sim M$,
we can rewrite (\ref{ccp}) as
$$
V_{min}\sim {3 \alpha K_+^4\over (4q-1)}\left\{
\left(\ln\left(K_+/\mu_+\right)+{1\over 3}\right)+
h^{8(q-1)/(2q-1)}\left(\ln\left(K_-/\mu_-\right)+{1\over 3}\right)
\right\}
$$
Here we are assuming that $\theta\neq 1-1/q$ (so that $\alpha\neq 0$),
since the alternative case was already discussed in the previous subsection.
For $q>1$, the first term dominates and the condition of a nearly
vanishing cosmological constant forces $K_+ \approx \mu_+ e^{-1/3}$.
A fine-tuning of $\Lambda_+$ will be necessary in order to
satisfy the condition (\ref{extremum}) for such value of $K_+$.
The hierarchy is given by
$$
h^2 \sim \left({\mu_+\over M}\right)^{2q-1} \sim
\exp\left[-(2q-1) \alpha^{-1} c_+ \right],
$$
where $\mu_+$ is given by (\ref{muy}).
Since the effective coupling $\alpha$ can be rather small,
a large hierarchy may be obtained even for moderate $q\gtrsim 1$.
A straightforward calculation shows that
at the minima of the effective potential (\ref{sumsup})
$
\partial^2_{\varphi_+}V = 12\alpha(1+2q)^{-2}
a_+^4 K_+^4 \varphi_+^{-2}$, and
$\partial^2_{\varphi_-}V \sim \alpha q^{-1}
a_-^4 K_-^4 \varphi_-^{-2}
$.
Hence, we find that the physical masses for the moduli fields
$\varphi_+$ and $\varphi_-$ which appear in (\ref{moder}) are given by
$$
m^2_+ \sim \alpha q^{-2} h^{12/(2q-1)}\ m_p^2\,,
\quad\quad m_-^2 \sim \alpha q^{-1} h^{2+ 4/(2q-1)} m_p^2.
$$
Associated with the eigenvalue $m_+$ there is a Brans-Dicke (BD) field
 \footnote{Here we are considering
the situation where the mass of $\varphi_-$ is much larger than the mass
of $\varphi_+$, and where the visible matter is on the negative tension
brane. In this case, since $y_-=const.$, visible matter is universally
coupled to the metric $g_{\mu\nu}$, and the BD parameter corresponding
to $\varphi_+$ can be read off from (\ref{moder}).  Alternatively,
it can be read off from (\ref{skin-}) after setting $y_-=const.$},
with BD parameter $\omega_{BD}=-3q/(1+2q)$.
Therefore, we must require $m_+ \gtrsim (mm)^{-1}$, which in turn
requires $q>2$. A stronger constraint on $q$ comes from the eigenvalue
$m_-$, since the corresponding field is coupled to ordinary matter
with $TeV$ strength. The mass of this field cannot be too far
below the $TeV$, otherwise it would have been seen in accelerators.
This requires $q$ to be rather large $q \gtrsim 10$.

\section{Conclusions}

We have studied a
class of warped brane-world compactifications, with a
power law warp factor of the form $a(y)=(y/y_0)^q$ and
a dilaton with profile $\phi\propto \ln(y/y_0)$. Here
$y$ is the proper distance in the extra dimension. In general,
there are two different moduli
$y_{\pm}$ corresponding to the location of the branes.
(in the RS limit, $q\to \infty$,
a combination of these moduli becomes pure gauge).

Classically, the moduli are massless, but they develop an effective
potential at one loop. We have presented methods
for calculating this effective potential, using both zeta function
and in dimensional regularization. The procedure used in the case
of zeta function regularization is a simpler version of the
one we introduced in Ref. \cite{gpt}.
Our treatment of dimensional regularization formalizes and extends the
approach adopted in Refs. \cite{gr,toms,flachitoms}. An important
point is that the divergent term to be subtracted from the
dimensionally regularized effective potential
is proportional to the the Seeley-de Witt coefficient $a_{5/2}$, given
in (\ref{betaf}). In the RS model, this coefficient
behaves much like a renormalization of the brane tension, but it
behaves very differently in the general case.

In general, the effective potential induced by massless
bulk fields with arbitrary curvature coupling is given by
(\ref{veffnonmin2}). In the limit when the branes are very
close to each other, it behaves like
$
V\propto a^4 |y_+-y_-|^{-4},
$
corresponding to the usual Casimir interaction in flat space.
Perhaps more interesting is the moduli dependence due to local operators
induced on the branes, which are the dominant terms in $V(y_+,y_-)$
when the branes are widely separated. Such operators break a scaling
symmetry of the classical action, which we discussed in Section II,
but nevertheless are needed in
order to cancel the divergences in the effective potential.
If we denote by
$K(y_i)=q/y_i$ the extrinsic curvature of the brane at the location
$y=y_i$ ($i=\pm$), a renormalization of the brane tension parameters $\sigma_\pm$ in the
classical action (\ref{general}) induces terms proportional to $a(y_i)^4 K_i$
in the effective potential. These terms scale like the rest of the classical
action under the global transformation (\ref{scaling1}-\ref{scaling2}).
On the other hand, the divergences in the effective
potential, proportional to the coefficient $a_{5/2}(P)$,
require world-sheet counterterms
which are proportional to $a(y_i)^4 K^4(y_i)$. These have the wrong scaling behaviour
[they simply do not change under (\ref{scaling1}-\ref{scaling2})] and hence they
act as stabilizers for the moduli.

In addition, there are terms proportional to
$a(y_i)^4 K^4(y_i) \phi(y_i)$. The coefficient in front
of the latter terms depends on the choice of the measure in the path integral.
Different choices are possible, which are related amongst each other by
dilaton dependent conformal transformations. Because of the conformal anomaly,
different choices are inequivalent,
but they are simply related by the
addition of local world-sheet operators to the action.
These are given by the r.h.s. of (\ref{onu}). Since the
scaling symmetry of the action is anomalous, the presence of these local
terms could perhaps have been guessed without the need of detailed calculation.
Since $\phi$ behaves logarithmically, these terms have the form of
Coleman-Weinberg type potentials for the moduli $y_i$, and they can also
act as stabilizers for the moduli.

To conclude, we find that worldsheet operators induced on the brane
at one loop easily stabilize the moduli in brane-world scenarios with warped
compactifications, and give them sizable masses. If the warp
factor is sufficiently steep, $q\gtrsim 10$, then
this stabilization naturally generates a large hierarchy, as in
the Randall-Sundrum model. In this case, the mass of the
lightest modulus is somewhat below the $TeV$ scale.
This feature is in common with the Goldberger and Wise mechanism
\cite{gw1} for the stabilization of the radion in the RS model. For
$q \lesssim 10$,
the stabilization is also possible, but if we also demand that the
hierarchy $h\sim 10^{-16}$ is generated geometrically, then the
resulting masses for the moduli would be too low.

\section*{Acknowledgements}

We are grateful to Klaus Kirsten and Alex Pomarol for useful discussions.
J.G. thanks the Tufts Cosmology Institute for hospitality while
part of this work was prepared. J.G. is partially
supported by the Templeton Foundation under grant COS 253, by CICYT
under grants AEN99-0766,AEN98-0431, by NATO under a collaborative
research grant and by the Yamada Foundation. T.T. thanks the IFAE,
where most of the present work was completed, for cordial hospitality.
T.T. is partially supported by the Monbukagakusho Grant-in-Aid No.~1270154,
and by the Yamada Foundation. O.P. is partially supported by
CICYT, under grant AEN99-0766 and by CIRIT under grant 1998FI 00198.
\appendix



\section{Another form of the moduli action}

In this Appendix, we present alternative forms of the moduli
action in terms of the metric induced on either brane. This
highlights the fact that the interactions with matter on the
branes are of Brans-Dicke type.

Changing variables in the background solution (\ref{metricform})
according to $y=y_-+R\Theta$, with $0<\Theta<\pi$, we can recast the
moduli as the radius of the
orbifold $R$ and the 'would be' distance $y_-$ from the
negative tension brane to the singularity.
Then the line element of our solution is
\begin{equation}
    ds^2=R^2\,d\Theta^2 + \left({y_-+R\,\Theta \over y_0}\right)^{2{\kk}}\,
    \eta_{\mu\nu}dx^\mu dx^\nu,
\end{equation}
with $R$ and $y_-$ arbitrary constants.

In order to take into account the gravitational degrees of freedom, we
promote $R$ and $y_-$ (and the induced 4d metric) to 4d 'moduli' fields depending
on $x^\mu$, so that the ansatz we shall consider takes the form
\begin{eqnarray}
    \label{ansatz}
    ds^2&=&R(x^\mu)^2\,d\Theta^2 + \big(1+r(x^\mu)\,\Theta \big)^{2{k}}\,
    g^{(-)}_{\mu\nu}dx^\mu dx^\nu,\\
    \phi&=&\sqrt{6{\kk}}\,\kappa_5^{1/2}\,\big[\ln\big(1+r(x^\mu)\,\Theta\big)+\ln
    \big(y_-(x^\mu)/y_0\big)\big]
\end{eqnarray}
Here we have absorbed $y_-(x^\mu)$ into the 4d metric $g^{(-)}_{\mu\nu}$, and
defined $r\equiv R/y_-$.  We insert this ansatz back to the action (\ref{general})
and integrate over $\Theta$ to get the kinetic terms of the 4d effective action.
In terms of the metric on the negative tension brane we find
\begin{eqnarray}
    \label{skin-}
    S_b={-2\over \kappa_5}\int d^4x \sqrt{g_{(-)}} y_-
    \Biggl\{&&\left({\lambda^{2{\kk}+1}-1\over2{\kk}+1}\right)  {\cal R}^{(-)}\cr
    &&+ 3 {\kk}\left({\lambda^{2{\kk}+1}-1\over2{\kk}+1}\right)
          {(\partial y_-)^2\over y_-^2}
          -3{\kk}\lambda^{2{\kk}+1}{(\partial \lambda)^2\over \lambda^2}
          \Biggr\},
\end{eqnarray}
where indices are contracted with the $g^{(-)}$ metric, ${\cal R}^{(-)}$
is the 4d
curvature scalar related to it, and we
defined
\begin{equation}
    \label{lambda}
    \lambda\equiv \Big({a(\pi)\over a(0)}\Big)^{1/{\kk}}={y_+\over y_-}=1+\pi\,r >1.
\end{equation}
Note that $\lambda$ and $y_-$  have diagonal kinetic terms,
with $\lambda$ depending on the hierarchy of scales between the branes only.

The contribution to the kinetic part of the effective
action from these moduli
as seen from the positive tension brane is obtained
from Eq. (\ref{skin-}) by
writing $g^{(-)}_{\mu\nu}$ 
as $\lambda^{-2{\kk}} g^{(+)}_{\mu\nu}$ (equivalently, we can change
$\lambda$ by
$1/\lambda$ and $y_-$ by $ y_+\equiv \lambda y_-$). After partial integration and a
global change of sign, we get
\begin{eqnarray}
    \label{skin+}
    S_b={-2\over \kappa_5}\int d^4x \sqrt{g_{(+)}} \,y_+
    \Biggl\{ &&\left({1-\lambda^{-(2{\kk}+1)}\over2{\kk}+1}\right)  {\cal R}^{(+)}\cr
    &&+3{\kk}\left({1-\lambda^{-(2{\kk}+1)}\over2{\kk}+1}\right)
    {(\partial y_+)^2\over y_+^2}
    +3{\kk}\lambda^{-(2{\kk}+1)}{(\partial \lambda)^2\over \lambda^2}
    \Biggr\},
\end{eqnarray}
with indices contracted with $g^{(+)}$.

It is easy to see that both $\lambda$ and $y_-$ or $y_+$ can be put in the form of
Brans--Dicke (BD) fields, with a generic action
\begin{equation}
    \label{bransdicke}
    -\int\,d^4x\,\sqrt{g}{1\over16\pi}\Big(\Phi_{BD}\,{\cal R}
    +{\omega_{BD}(\Phi_{BD})\over \Phi_{BD}} (\partial \Phi_{BD})^2 \Big)\; .
\end{equation}
Ignoring the global $\lambda$-dependent factor, $y_-$ and $y_+$ are pure
Brans--Dicke field up to a rescaling, with Brans--Dicke parameter
$\omega^{y_\pm}_{BD}=3{\kk}$.
In the case of $\lambda$, again ignoring the global $y_-$ (or $y_+$)
factor, the
Brans--Dicke parameter are
\begin{equation}
    \label{omegalambdabd}
    \omega^{\lambda~(\pm)}_{BD}(\lambda)={3{\kk}\over2{\kk}+1}
                         \big( \lambda^{\pm(2{\kk}+1)}-1 \big).
\end{equation}
From the positive tension brane, and for ${\kk}\sim1$, the observational limit,
$\omega_{BD}>3000$ \cite{will}, is fulfilled with $\lambda\sim15$.  This
requires an interbrane proper distance $\pi R$ at least about 1 order
of magnitude
larger than the distance from the singularity at $y=0$ to the negative tension
brane, at $y=y_-$.
In the negative tension brane, we find that the BD parameter is negative
but greater than $-3/2$ for any $\lambda$ or ${\kk}$, as happened in \cite{gt}.
We reproduce the results of \cite{chiba} in the limit of ${\kk}\to\infty$.

\section{Dimensional regularization versus zeta function regularization}
\label{app:dimreg}

\subsection{Zeta function regularization}

Let us rederive the main result of Section V using zeta function
regularization.
For simplicity, we shall restrict attention to the case $\theta=1$.
In this case, we have explicitly calculated the integral (\ref{integ})
along the conformal path using both methods described in Section IV.
This exercise leads to the
result
\begin{eqnarray}
    V= V_0 - \frac{1}{(4 \pi)^2}\Biggl[-\beta_4
    \left({1\over z_-^4}\ln  (z_{-}/z_0)^\beta +{1\over z_+^4}
    \ln (z_{+}/z_0)^\beta \right)  +\hat\alpha
    \left({1\over z_-^4}+{1\over z_+^4}\right)    \Biggr],
    \label{int1}
\end{eqnarray}
where
$
\hat\alpha= (\beta/3072)\{144 + \beta \big(784 - 1692 \beta + 335 \beta^2 -
192 (3 \beta-14) ( 3 \beta-2) \xi\big)\}
$.
The zeta function associated with the operator $P_0$ is given by
\begin{equation}
 \zeta_{0}(s)={\cal A}\int {d^{4} k\over (2\pi)^{4}}
        \sum_{i}\left({k^2+m_n^2\over \mu^2}\right)^{-s}.
\label{yjw}
\end{equation}
Performing the momentum integrals in (\ref{yjw}), we have
\begin{equation}
\zeta_{0}(s) = {\cal A} {\mu^{2 s} z_{-}^{2s-4}
             \Gamma(s-2)
             \over (4\pi)^{2}\Gamma(s)}\tilde \zeta(2s-4),
\label{z0}
\end{equation}

Substituting (\ref{z0}) into (\ref{vtheta}), we have
\begin{equation}
 V_0(z_+,z_-)  =  -{1\over
        2(4\pi)^2 z_-^{4}}\left[
       \left\{\ln(\mu z_-)+{3\over 4}
         \right\}\tilde\zeta(-4)
          +\tilde\zeta'(-4)\right],
\label{V0}
\end{equation}
and from (\ref{tildezeta}) we obtain
\begin{equation}
    V_0(z_+,z_-)={1\over (4\pi)^2}
    \left[{{\cal I}_K \over z_<^4}+ {{\cal I}_I\over z_>^4} +
    {{\cal V}(\tau)\over z_>^4}\right]
    + {\beta_4 \over (4\pi)^{2}}
    \left[{1\over z_+^4} \ln(\mu z_+) +{1\over z_-^4 }\ln(\mu z_-)\right].
    \label{flatstresult}
\end{equation}

Substituting in (\ref{int1}) we recover Eq. (\ref{veffnonmin2}) up to finite
renormalization of $\mu$.

\subsection{$V_0$ in dimensional regularization}
\label{sec:dimreg}

We shall now reproduce the result for $V_0$
by using dimensional regularization. Again, this is a
redundant exercise:
the calculation of an effective potential (be it $V$ or $V_0$)
will give the same answer whether it is done in dimensional or
in zeta function regularization. Nevertheless, it is
interesting to do it explicitly since this calculation is closest
in spirit to the standard flat space calculations in four dimensions.

Adding up the dimensionally regularized effective potential per
comoving 4-volume due to all KK modes, we have
\begin{equation}
    \label{casimir}
    V_0^{\text{reg}} = {1\over 2}\mu^\epsilon \sum_n\int {d^{4-\epsilon}
      k\over
      (2\pi)^{4-\epsilon}}\log\left({k^2 + m_n^2 \over\mu^2}\right).
\end{equation}
Performing the momentum integration for each mode, we obtain
\begin{equation}
    V_0^{\text{reg}}=
    - {1\over 2(4 \pi)^2}(4\pi\mu^2)^{\epsilon/2}
    {1\over z_-^{4-\epsilon}}
   \Gamma(-2+\epsilon/2) \sum_n \tilde m_n^{4-\epsilon},
\end{equation}
where we used $\tilde m_n$ defined in (\ref{tildem}).
This regularized expression for the effective potential
is finite when the real part of $\epsilon$ is sufficiently large.
Performing analytical continuation in $\epsilon$,
the summation over KK modes $\sum_n \tilde
m_n^{4-\epsilon}$ can be identified with
the zeta function $\tilde\zeta(-4+\epsilon)$.
The pole part proportional to $1/\epsilon$ is identified with
\begin{equation}
    \label{4dcountt}
    V_0^{\text{div}}
    =-{1\over\epsilon}{1\over2(4\pi)^2 z_-^4}\tilde\zeta(-4).
\end{equation}
Subtracting this divergent part, we get the renormalized expression
for the effective potential as
\begin{equation}
    \label{4dfiniteresult}
    V_0=V^{\text{reg}}-V^{\text{div}}=-\lim_{\epsilon\to0}
    {1\over 2(4 \pi)^2}\left[
        {(4\pi\mu^2)^{\epsilon/2} \over z_-^{4-\epsilon}}
        \left({1\over\epsilon}+{3\over4}-{\gamma\over2} \right)
        \tilde\zeta(-4+\epsilon)
        -{1\over\epsilon} \tilde\zeta(-4)\right].
\end{equation}
Consequently, we find that the dimensional regularization
method reproduces the previous result (\ref{flatstresult}).

Here, one remark is in order. In a usual 4-dimensional problem,
the divergent part is given by the
Seeley-De\,Witt coefficient $a_{2}$.
In the present case, since the background
is 4-dimensional flat space, $a_{2}$ consists of
only one term proportional to $m_n^4$ for each KK mode.
Subtraction of this counter term for each mode
leads to the expression $\sum_n m_n^4 \log m_n$, which is still
divergent. This is not a surprising fact because
the problem is essentially 5-dimensional as it is indicated by
the existence of the infinite tower of the KK modes.
Usually, this point is bypassed in the literature by evaluating
the divergent sum with the help of the generalized zeta function.

\subsection{Alternative regularization}

In section V, we were interested in the class of conformally related
operators $P_\theta$, and therefore it was important to do a dimensional
extension of the spacetime such that all geometries labeled by
$\theta$ would be conformally flat. In other words, we added
dimensions "parallel" to the brane, whose "size" was also affected by
the warp factor. It should be stressed that this was done for
computational convenience, since $V^D$ is independent of the
parameter $\theta$ only in this regularization.
Putting aside computational considerations, nothing prevents us
to extend the spacetime in any way we please, and the results
should still be the same. To illustrate this point,
let us consider an alternative dimensional extension of our 5-dimensional
curved space ${\cal M}$ to a simple direct product space given by
$\Bbb R^{-\epsilon}\times{\cal M}$.
The dimensional regularization is done by an analytic continuation
of the number of added dimensions $\epsilon$,
while the manifold ${\cal M}$ is kept unchanged.

Since this is a direct product space,
the eigenvalues of the ($5-\epsilon$)-dimensional d'Alembertian
are given by a simple summation of
those in each space, ${\cal M}$ and $\Bbb R^{-\epsilon}$,
$$
\lambda_{(5-\epsilon)}=k_{\omega}^2+\lambda.
$$
Then, the dimensionally regularized effective potential
per unit comoving volume is given by
\begin{eqnarray}
    V^{\text{reg}}
    ={1\over 2} \mu^\epsilon
    \sum_{\lambda} \int
   {d^{-\epsilon} k_{\omega}\over (2\pi)^{-\epsilon}}
   \log\left({k_\omega^2+\lambda\over \mu^2}\right).
\label{5dV}
\end{eqnarray}
This quantity is evaluated by introducing the function
\begin{equation}
    \label{5dzeta}
    {\Upsilon}(s)
    =\sum_{\lambda} \int {d^{-\epsilon} k_{\omega}\over (2\pi)^{-\epsilon}}
    \left({k_{\omega}^2+\lambda\over \mu^2}\right)^{-s} \nonumber\\
    = (1/4\pi)^{-\epsilon/2}{\Gamma(s+\epsilon/2)\over
    \Gamma(s)}\zeta_1(s+\epsilon/2),
\end{equation}
where $\zeta_1$ is the zeta function defined in (\ref{zepsilon}) with
$\theta=1$.
Then, using (\ref{5dzeta}), we can rewrite (\ref{5dV}) as
\begin{equation}
    \label{5dregeffpot}
    V^{\text{reg}}=
     -{1\over 2 {\cal A}} {\Upsilon'(0)}
     =
    -{1\over2{\cal A}}(1/4\pi)^{-\epsilon/2}
    \Gamma(\epsilon/2) \zeta_1 (\epsilon/2).
\end{equation}
The above expression contains the object $\zeta_1$, which
we have encountered in zeta function regularization.
However, it should be noted that now the regularization
parameter is not $s$, the argument of the function $\Upsilon$,
but the dimension of the product space $\epsilon$.
Therefore, even after we take the limit $s\to 0$,
$V^{\text{reg}}$ still diverges as $1/\epsilon$.

The divergent piece
in the dimensional regularization in the $D$ dimensional problem
is the Seeley-De\,Witt coefficient $a_{D/2}$.
From (\ref{zepsilon}) and (\ref{sdw}), it is easy to relate
this divergent piece with the value of $\zeta_1(0)$ as\footnote{
Strictly speaking $a^D_{5/2}$ has to be evaluated in the
regularized $(5+\epsilon)$-dimensional space.
However, since the added $\epsilon$ dimensions
are trivial, it is identical to $a_{5/2}(P_1)$.}
\begin{equation}
    \label{5ddiv0}
    V^{\text{div}}={-a_{5/2}(P_1) \over \epsilon{\cal A} }
       = {-1\over \epsilon {\cal A}} \zeta_1(0).
\end{equation}
Subtracting this divergent piece from (\ref{5dregeffpot}), we
obtain the renormalized value
\begin{equation}
    \label{v1}
    V=V^{\text{reg}}-V^{\text{div}}
    =-{1\over2{\cal A}}\Bigl(\zeta'_1 (0) -
    \{\ln\big(1/4\pi\big)+\gamma \}
             \zeta_1 (0)\Bigr),
\end{equation}
where $\gamma$ is Euler's gamma.
This coincides with Eq. (\ref{vtheta}) for $\theta=1$
up to a redefinition of $\mu$
(note that with our conventions, $\zeta'_1 (0)$ depends on $\mu$.)

\section{Definition of the constants ${\cal I}_I$ and ${\cal I}_K$}

The result (\ref{skippy}) involves the two constants ${\cal I}_I$ and
${\cal I}_I$. Although the actual value of this constants seems to be
irrelevant, since they can be shifted by finite renormalization of
worldsheet operators, we will nevertheless for the sake of
completeness give their definition in terms of the parameters of the model.

We can express the asymptotic expansion of the combinations of
modified Bessel functions used in subsection \ref{sec:nonmin} as
\begin{eqnarray}
  &&I_{\nu}(\rho)\sim {e^{\rho}\over \sqrt{2\pi \rho}}
\sum_{r=0}^{\infty}{\delta^{(I)}_r \over \rho^r}
             +O(e^{-\rho}),\cr
  &&K_{\nu}(\rho)\sim \sqrt{\pi\over \rho} e^{-\rho}
\sum_{r=0}^{\infty}{\delta^{(K)}_r \over \rho^r}.
\end{eqnarray}
Since both $\delta^{(I)}_0=\delta^{(K)}_0=1$, and the obsevration that
$\beta^{(K)}_4=\beta^{(I)}_4 \equiv\beta_4$, we can define coefficients
$\beta_r$
for both $K_\nu$ and $I_\nu$ as
\begin{equation}
    \ln \left( \sum_{r=0}^{\infty}{\delta_r \over \rho^r} \right)=
    \sum_{r=1}^{\infty} {\beta_r \over \rho^r}.
\label{betas}
\end{equation}
Defining the functions
\begin{eqnarray}
    R^{(K)} (\rho)&=&\sum_{r=1}^{3} {\beta^{(K)}_r\over \rho^r}
    +{\beta_{4}\over \rho^{4}} e^{-1/\rho}, \nonumber \\
    R^{(I)}(\rho)&=&\sum_{r=1}^{3} {\beta^{(I)}_r\over \rho^r}
    +{\beta_{4}\over \rho^{4}} e^{-1/\rho},
\end{eqnarray}
the coefficients ${\cal I}_I$ and ${\cal I}_K$ are defined as
\begin{eqnarray}
    {\cal I}_I&=&-\gamma \beta_{4}+\int_{0}^{\infty} d\rho\, \rho^{3}
    \left(\ln\left[\sqrt{2\pi\rho}\; e^{-\rho}I_{\nu}(\rho)\right]
        - R^I(\rho)\right),  \nonumber \\
    {\cal I}_K&=&-\gamma \beta_{4}+\int_{0}^{\infty} d\rho\, \rho^{3}
    \left(\ln\left[\sqrt{2\rho/\pi}\, e^{\rho}K_{\nu}(\rho)\right]
        - R^K(\rho)\right).
\end{eqnarray}

\section{Asymptotic form of ${\cal V}(\tau)$}

The behaviour of ${\cal V}(\tau)$ defined in (\ref{calvo})
for $\tau\ll 1$ is given by
\begin{equation}
{\cal V}(\tau)=
-{2  \over \nu\Gamma(\nu)^2}\left({\tau\over
2}\right)^{2\nu}\int_0^\infty dt\,t^{2\nu+3} {K_{\nu} (t)\over
I_{\nu}(t)}+{\cal O}\left(\tau^4 \ln \tau\right).
\end{equation}
This corresponds to a large separation between branes. In this limit
the integral is generically negligible compared
with the logarithmic terms in (\ref{veffnonmin2}).

The limit of small separation between branes corresponds to $1-\tau\ll 1$.
In this limit, the integral can be approximated by taking the arguments
of the Bessel functions to be large. Using the asymptotic expansion
\begin{equation}
{I_{\nu}(\tau\rho) K_{\nu}(\rho)\over
K_{\nu}(\tau\rho) I_{\nu}(\rho)} \sim e^{-2(1-\tau)\rho},
\end{equation}
we have
\begin{eqnarray}
{\cal V} & \approx & \int_0^{\infty}d\rho \rho^3\ln\left(
   1-e^{-2(1-\tau)\rho}\right) \cr
   & = & -{1\over 2^6 (1-\tau)^4}\int_0^{\infty}
       dx {x^4\over e^x -1} =-{3\zeta(5) \over 8 (1-\tau)^4}.
\end{eqnarray}
In the second equality, we performed integration by parts and
a change of variable. Here, $\zeta$ is the usual Riemann's zeta function
$\zeta(5)=1.03693\cdots$. Using the relation
\begin{equation}
 \zeta'(-4)= {3\over 4\pi^4}\zeta(5),
\label{zetap}
\end{equation}
and substituting in (\ref{veffnonmin2}), we find that in the limit
of small brane separation, the effective potential reduces to the
one we had found in the massless conformally coupled case, given
in equation (\ref{ve1}):
\begin{equation}
    V(z_+,z_-) \sim - {A \over |z_+ - z_-|^4}.
\end{equation}
Equation (\ref{zetap}) is a particular case of a more general
formula
$$
 \zeta'(2n)={(2n)!\over 2^{2n+1}\pi^{2n}}\zeta(2n+1),
$$
valid for positive integer $n$. This can be derived
from the perhaps better known relation \cite{takahirotable}
$$
 \zeta(1-z)=2^{1-z}\pi^{-z} \zeta(z)\Gamma(z)\cos(\pi z/2),
$$
by setting $z=2n+1$ after differentiation of both sides of the
equation with respect to $z$.

\end{document}